\documentclass[traditabstract]{aa} 
\usepackage{graphicx}

\usepackage{lscape}
\usepackage{amssymb}
\usepackage{amsmath}

\usepackage{txfonts}

\usepackage{booktabs}
\usepackage{subfigure}
\usepackage{placeins}
\usepackage{verbatim}
\usepackage{adjustbox}
\usepackage[toc,page]{appendix} 
\usepackage{tabularx}
\usepackage{lscape}
\usepackage{textcomp}
\usepackage{gensymb}
\usepackage{hyperref}
\usepackage{multicol}
\usepackage{tikz}
\usepackage{caption}
\usepackage{longtable}
\usepackage{xspace}

\usepackage{natbib}
\usepackage{multirow}

\usepackage{color}
\usepackage{xtab}
\usepackage{orcidlink}

\newcommand{\bd}{\begin{displaymath}}
\newcommand{\ed}{\end{displaymath}}
\newcommand{\be}{\begin{equation}}
\newcommand{\ee}{\end{equation}}
\newcommand{\beaa}{\begin{eqnarray*}}
\newcommand{\eeaa}{\end{eqnarray*}}
\newcommand{\bea}{\begin{eqnarray}}
\newcommand{\eea}{\end{eqnarray}}

\def\GLEE{\textsc{Glee}\xspace}

\def\gradeB{503\xspace}
\def\gradeA{95\xspace}
\def\rediscoveries{506\xspace}
\def\newdiscoveries{92\xspace}

\def\gleeautopy{\textit{glee}$ \_ $\textit{auto.py} \xspace}

\defcitealias{canameras21b}{C21}
\defcitealias{shu22}{S22}
\defcitealias{schuldt25a}{S25}
\defcitealias{schuldt23a}{S23a}
\defcitealias{canameras24}{C24}

\usepackage[modulo, switch]{lineno}

\begin{document}

   \title{HOLISMOKES XVI. Lens search in HSC-PDR3 with a neural network committee and post-processing for false-positive removal}

  \titlerunning{HOLISMOKES XVI: Lens search in HSC-PDR3}

   \author{{S. Schuldt}\inst{1}\inst{,2\orcidlink{0000-0003-2497-6334}}
                        \and
    {R. Cañameras}\inst{3\orcidlink{0000-0002-2468-5169}}
    \and
    {Y.~Shu}\inst{4\orcidlink{0000-0002-9063-698X}}
    \and
    {I.~T.~Andika}\inst{5}\inst{,6\orcidlink{0000-0001-6102-9526}}
    \and
    S.~Bag\inst{5}\inst{,6\orcidlink{0000-0003-0141-606X}}
    \and
    {C.~Grillo}\inst{1}\inst{,2\orcidlink{0000-0002-5926-7143}}
    \and
    {A. Melo}\inst{6}\inst{,5\orcidlink{0000-0002-6449-3970}}
    \and \\
    {S.~H.~Suyu}\inst{5}\inst{,6\orcidlink{0000-0001-5568-6052}}
    \and
    {S.~Taubenberger}\inst{5}\inst{,6\orcidlink{0000-0002-4265-1958}}
                        }

        \institute{Dipartimento di Fisica, Universit\`a  degli Studi di Milano, via Celoria 16, I-20133 Milano, Italy\\
        e-mail: \href{mailto:stefan.schuldt@unimi.it}{\tt stefan.schuldt@unimi.it}
        \and
        INAF -- IASF Milano, via A. Corti 12, I-20133 Milano, Italy
        \and
        Aix-Marseille Université, CNRS, CNES, LAM, Marseille, France
        \and
        Purple Mountain Observatory, No. 10 Yuanhua Road, Nanjing, Jiangsu, 210033, People’s Republic of China
        \and
        Technical University of Munich, TUM School of Natural Sciences, Physics Department,  James-Franck-Stra{\ss}e 1, 85748 Garching, Germany
        \and
        Max-Planck-Institut f{\"u}r Astrophysik, Karl-Schwarzschild Stra{\ss}e 1, 85748 Garching, Germany
         }

   \date{Received --; accepted --}

 
  \abstract{
We have carried out a systematic search for galaxy-scale lenses exploiting multiband imaging data from the third public data release of the Hyper Suprime-Cam (HSC) survey with the focus on false-positive removal, after applying deep learning classifiers to all $\sim$110 million sources with an $i$-Kron radius above 0\farcs8. To improve the performance, we tested the combination of multiple networks from 
our previous lens search projects and found the best performance by averaging the scores from five of our networks. Although this ensemble network leads already to a false-positive rate (FPR) of $\sim$ 0.01\% at a true-positive rate (TPR) of 75\% on known real lenses, we have elaborated techniques to further clean the network candidate list before visual inspection. In detail, we tested the rejection using SExtractor and the modeling network from HOLISMOKES IX, which resulted together in a candidate rejection of 29\% without lowering the TPR. After the initial visual inspection stage to remove obvious non-lenses, 3,408 lens candidates of the $\sim$110 million parent sample remained. We carried out a comprehensive multistage visual inspection involving eight individuals and identified finally \gradeA grade A (average grade $G \geq 2.5$) and \gradeB grade B ($2.5 > G \geq 1.5$) lens candidates, including \newdiscoveries discoveries showing clear lensing features that are reported for the first time. This inspection also incorporated a novel environmental characterization using histograms of photometric redshifts. We publicly release the average grades, mass model predictions, and environment characterization of all visually inspected candidates, while including references for previously discovered systems, which makes this catalog one of the largest compilation of known lenses. The results demonstrate that (1) the combination of multiple networks enhances the selection performance and (2) both automated masking tools as well as modeling networks, which can be easily applied to hundreds of thousands of network candidates expected in the near future of wide-field imaging surveys, help reduce the number of false positives, which has been the main limitation in lens searches to date.}


   \maketitle

\section{Introduction}
\label{sec:intro}

Strong gravitational lensing has emerged in recent decades as a powerful tool to probe galaxy evolution and cosmology. It allows us to obtain in a very precise way the total mass (i.e., baryonic and dark matter (DM)) of the galaxy or galaxy cluster acting as the lens \citep[e.g.,][]{bolton08, shu17, caminha19} and, by assuming that mass follows light, we can disentangle the mass components and obtain unique insights into the DM distribution \citep[e.g.,][]{schuldt19, shajib21, wang22} or DM substructure \citep[e.g.,][]{ertl24_HE0230, lange25, enzi25}. Thanks to the lensing magnification, strong lensing also allows us to study high-redshift sources not visible otherwise \citep[e.g.,][]{shu18, vanzella21, mestric22, stiavelli23, morishita24}. 

In the case of a time-variable background object, such as a supernova (SN) or a quasar, time delays can be measured \citep[e.g.,][]{courbin18, millon20b} and exploited for competitive measurements of the value of the Hubble constant, $H_0$ \citep{refsdal64}. Given the rarity of SNe and strong lensing events, this time-delay cosmography (TDC) technique was mostly carried out with quasars \citep[e.g.,][]{wong20, acebron22_SDSS2222, shajib22, acebron24_extendedmodel}. To date only three strongly lensed SNe are known with time delays usable for a precise measurement of $H_0$: SN Refsdal lensed by the cluster MACS J1149.5$-$2223 \citep[e.g.,][]{grillo18, kelly23, grillo24}, SN H0pe strongly lensed by the cluster PLCK~G165.7$+$67.0 \citep{frye24, pascale25}, and SN Encore with SN Requiem, both lensed by the same cluster, MACS J0138$-$2155 \citep[][Pierel et al. in prep.]{rodney21, pierel24, granata25_encore, ertl25_encore}. To prepare a systematic search with current and upcoming wide-field imaging surveys, we initiated the Highly Optimized Lensing Investigations of Supernovae, Microlensing Objects, and Kinematics of Ellipticals and Spirals \citep[HOLISMOKES][]{suyu20} program. As a precursor to the Legacy Survey of Space and Time (LSST) of the Vera C. \textit{Rubin} Observatory \citep{ivezic08, ivezic19}, we are currently exploiting data from the Hyper Suprime-Cam (HSC), which are expected to be very similar. 

In \citet[][hereafter \citetalias{canameras21b}]{canameras21b} we presented a residual neural network to search broadly for any static lens, and complemented this in \citet[][hereafter \citetalias{shu22}]{shu22} with deflectors at relatively high redshift ($z_\text{d} \geq 0.6$) to better cover the whole redshift range. Both projects relied on a single convolutional neural network (CNN) and targeted HSC images of the second public data release (PDR2). In this work, we combine multiple networks into an ensemble network, relax our restrictions on filter coverage to increase the observed sky area (requesting $gri$ bands instead of $grizy$), and consider data from PDR3 with a slightly deeper and larger footprint. 

While the resulting sample of network candidates from HSC is small enough for visual inspection, this will not be the case for LSST, \textit{Euclid} \citep{laureijs11}, \textit{Roman} \citep{spergel15_roman}, and the Chinese Space Station Telescope \citep{gong19}, delivering more than a billion images, even with a citizen science approach \citep[e.g.,][]{holloway24, holloway25_Q1, walmsley25}. Consequently, we urgently need further automated ways to lower the false-positive rate (FPR) before an unavoidable visual inspection.

In this paper, we show our deep learning ensemble network and, following \citetalias{canameras21b}, apply it to cutouts of any object with an $i$-Kron radius above 0\farcs8 that passes standardized HSC image quality flags (see \citetalias{canameras21b} for details). We explore two different approaches to reject false positives: (1) we run SExtractor to reject images with artifacts or without a real astrophysical source in the cutout, followed by a exclusion through HSC-pixel flags, and (2), we run the residual neural network from \citet[][hereafter \citetalias{schuldt23a}]{schuldt23a} to reject systems based on the mass model predictions. Specifically, Sect.~\ref{sec:network} presents the tested network committee and their performances on known real lenses. We then describe the two approaches of contaminant rejection in Sect.~\ref{sec:cleaning}, before we carry out a visual inspection as described in Sect.~\ref{sec:visualinspection} and present the newly discovered lens candidates. Finally, we give a summary and conclusion in Sect.~\ref{sec:conclusion}. We exploit trained networks from \citetalias{canameras21b}, \citetalias{shu22}, and \citet[][hereafter \citetalias{canameras24}]{canameras24}, and consequently also adopt a flat $\Lambda$CDM cosmology with $\Omega_\text{m} = 1-\Omega_\Lambda$ = 0.32 \citep{planck20} and $H_0 = 72\, \text{km s}^{-1} \text{Mpc}^{-1}$ \citep{bonvin17}.

\FloatBarrier
\section{Inference with network committees}
\label{sec:network}

We followed the approach described in \citetalias{canameras21b} to classify the $\sim$110 million galaxies from HSC Wide PDR3 with an $i$-band Kron radius of $\geq$0\farcs8 and optimize the contamination rate of the candidate strong lens sample. Based on test sets drawn from HSC Wide PDR2 images and designed to closely match a real classification setup, \citet{canameras24} show that the purity and overall classification performance are significantly improved with committees of multiple neural networks \citep[see also e.g.,][]{andika23}. The highest gain is obtained when combining networks trained on different ground-truth datasets, with different prescriptions for the parameters' distributions over the mock lenses used as positive examples. Taking the average or multiplication of output scores from networks that have little overlap in false positives due to their internal representations is best for improving the true-positive rate (TPR), defined as true-positives over all considered positives, at low FPR.

We investigated several combinations of neural networks chosen among the best performing ResNet and classical CNNs from \citetalias{canameras21b}, \citetalias{shu22}, and \citetalias{canameras24}. All considered networks reached on their own a TPR of 40-60\% at a FPR of 0.01\% using 70,910 non-lens images from the COSMOS field. We refer to the original publications for more details such as their receiver operating characteristic curves. 

After classifying the $\sim$110 million HSC PDR3 image cutouts with these networks, we compared the TPR over three independent sets of confirmed and candidate strong lenses as a function of the total number of lens candidates predicted by the committee. This allowed us to find the optimal committee that maximizes the TPR at a given output sample size (i.e., corresponding to a fixed human inspection time). The first set of test lenses (set 1) includes the 1,249 galaxy-scale grade A or B candidates from the SuGOHI papers\footnote{see \url{https://www-utap.phys.s.u-tokyo.ac.jp/~oguri/sugohi/}}. The second set (set 2) focuses on 201 galaxy-scale grade A from SuGOHI, which are also included in set 1. The third set (set 3) corresponds to the cleaned set of 178 galaxy-scale grade A or B lens candidates used in \citetalias{canameras21b} \citetalias{canameras24}, which excludes visually identified systems with multiple lens galaxies, or significant perturbation from the environment. 

\begin{figure}[t!]
\centering
\includegraphics[width=.5\textwidth]{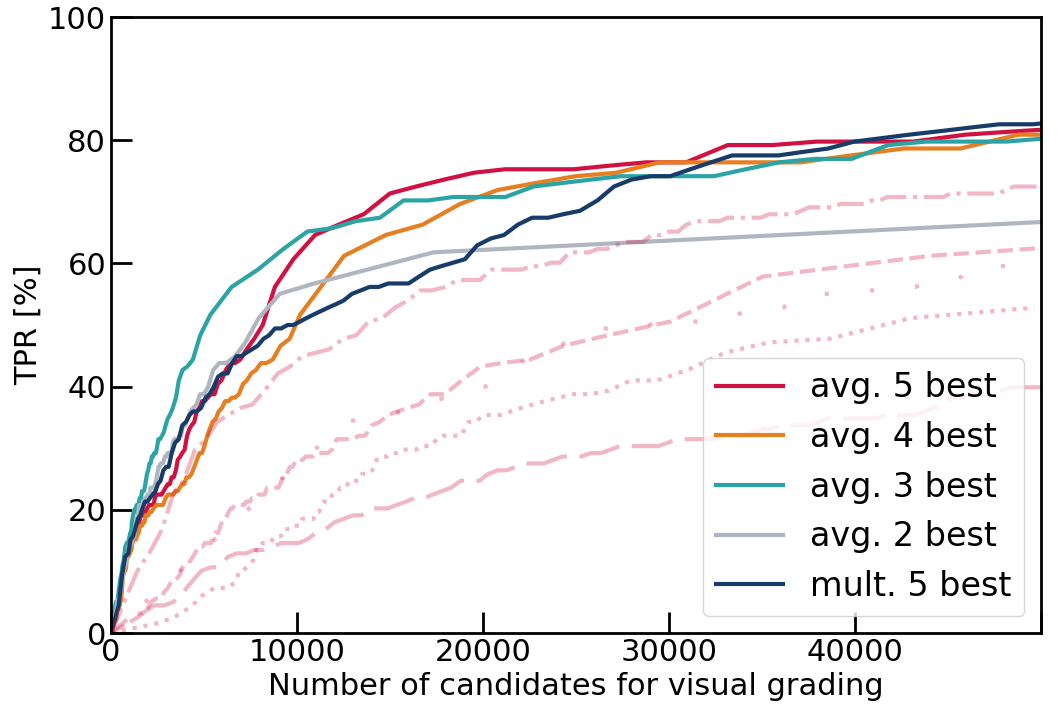}
\caption{True-positive rate (TPR), giving the ratio between true-positives and all positives, of different network committees (solid lines) measured on set 3, as a function of the number of lens candidates selected by each committee over the parent sample of 110 million sources. We consider here five different networks, whose individual curves are shown in light purple of various line styles. As examples, we show the performance of the ResNet trained on sets L4 and N1 from \citetalias{canameras24} (dash-dotted) and Classifier-1 from \citetalias{shu22} (long-dotted). The solid gray line shows the performance when averaging the scores from these two networks, the green one when additionally including the ResNet trained on L7 and N1 (long dashed), the orange one when also including the baseline ResNet from \citetalias{canameras24} (dotted), and red when additionally using the network from \citetalias{canameras21b} (dashed). We further show the TPR evolution with a solid dark blue line when multiplying the network scores from all five networks instead of averaging.}
\label{fig:roc_test}
\end{figure}

\begin{figure}[t!]
\centering
\includegraphics[width=.5\textwidth]{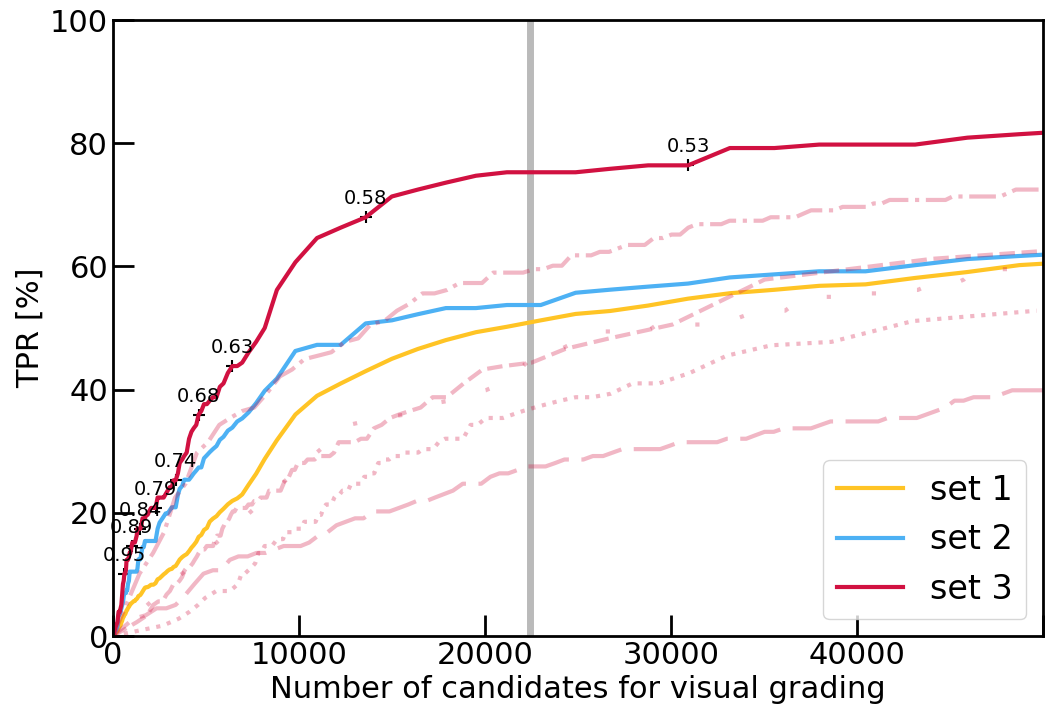}
\caption{Evolution of the TPR, as measured over the three test sets of candidate and confirmed strong-lenses, as a function of the number of candidates selected by the neural networks. Solid curves show the results for the final committee of five networks trained on different ground-truth datasets. Light purple curves show the lower TPR obtained on set 3 for the five individual networks included in the committee: the ResNet from \citetalias{canameras21b} (dashed curve), Classifier-1 from \citetalias{shu22} (long-dotted curve), the baseline ResNet from \citetalias{canameras24} (dotted curve), a network trained on mocks with a natural $\theta_{\rm E}$ distribution (dash-dotted curve), and a ResNet trained on balanced fraction of doubles and quads (long-dashed curve). The vertical gray line marks the final score threshold applied to select the list of strong-lens network candidates. Examples of score thresholds are marked for the evaluation of the best committee on set 3.}
\label{fig:roc}
\end{figure}

We investigated various combinations of different networks by taking the average or the product of their individual network scores. As examples, the performance of combinations with two, three, four, and five networks are shown in Fig.~\ref{fig:roc_test}. The best committee that maximizes the TPR in all three test sets at a fixed number of candidates was obtained by averaging the scores of five individual networks: the ResNet from \citetalias{canameras21b}, Classifier-1 from \citetalias{shu22}, and three additional ResNets from \citetalias{canameras24}, as is illustrated in Fig. \ref{fig:roc}. We adopted a threshold on the average score of 0.55, which corresponds to 22,393 strong-lens candidates, and a TPR of 51\%, 54\%, and 75\% over test set 1, 2, and 3, respectively. Since the TPR curves reach a plateau for all three of the test sets considered, decreasing the score threshold to include 50,000 additional candidates would have improved the TPR by only $\simeq$4–5\%, which does not justify nearly tripling the human inspection time\footnote{Using a threshold of $p=0.55$, we obtained around 20,000 network candidates. Therefore, 50,000 additional systems would increase the system by a factor of $\sim2.5$, leading to the specified cut on the network score. Since we recorded the annotation time during our visual inspection described in Sect.~\ref{sec:visualinspection}, we also estimated -- a posteriori -- the actual inspection time that would have been necessary for the additional sample. Here we obtained a factor $\sim3$ of the additional time, and possibly $\sim30$ additional lenses.}. Having various test sets was important, so that we could check that the plateau in TPR is reached irrespective of the exact setup of the test set. Since set 3 best represents the galaxy-galaxy scale lenses we target with this network committee (e.g., set 3 is cleaned from group and cluster lenses), the network committee shows the best performance on this set as expected. Finding about 20,000 candidates among 110 million sources from the parent sample is consistent with the FPR in the range of 0.01--0.03\% predicted in \citetalias{canameras24} for network committees, at a TPR of 75\% evaluated among set 3.

The five networks are all from the same type of architecture and include residual neural network blocks \citep{he16}. Specifically, the Classifier-1 from \citetalias{shu22} is based on the CMU DeepLens package from \citet{lanusse18}, while the other four networks are adapted from the ResNet-18 architecture. The networks were, however, trained on different ground-truth datasets. We introduce the general properties of the training sets and refer the reader to \citetalias{canameras21b}, \citetalias{shu22}, and \citetalias{canameras24} for further details.

For all five networks, the realistic mock lenses used as positive examples were simulated with the pipeline described in \citet{schuldt21b,schuldt23b}. Briefly, the pipeline paints lensed arcs on HSC images of massive luminous red galaxies (LRGs) using singular isothermal ellipsoid (SIE) lens mass profiles, and SIE parameter values inferred from SDSS spectroscopic redshifts and velocity-dispersion measurements. After the high-redshift background sources are drawn from the {\it Hubble} Ultra-Deep Field \citep[HUDF,][]{inami17}, mock lensed arcs are computed with \GLEE \citep{suyu10a_GLEE,suyu12}, and convolved with the HSC point spread function, before coaddition with the lens galaxy cutout. Positive examples used to train the ResNet from \citetalias{canameras21b} and Classifier-1 from \citetalias{shu22} were drawn from the same parent set of mocks, with (i) a nearly uniform Einstein radius distribution between 0\farcs75 and 2\farcs5, (ii) a boosted fraction of lens galaxies at $z > 0.7$ with respect to the parent SDSS sample, (iii) lensed images that have $\mu \geq 5$, and (iv) a minimal ratio of $ {\rm SNR}_{\rm bkg,min} = 5$ between the brightest arc pixel and the local sky background over the lens LRG cutout. The other three high-performing networks from \citetalias{canameras24} that are part of the committee were trained on mocks produced with a similar procedure, but with (i) a boosted fraction of red HUDF sources, (ii) a natural Einstein radius distribution between 0\farcs75 and 2\farcs5 instead of an uniform distribution, and (iii) no lower limit on $\mu$ and a balanced fraction of double and quad configurations. This corresponds to sets labeled L1, L4, and L6 in \citetalias{canameras24}. 

In terms of negative examples, four out of the five networks include a mix of 33\% spirals, 27\% isolated LRGs without arcs, 6\% groups, and 33\% random galaxies over the HSC footprint, as is defined for set N1 in \citetalias{canameras24}. The fifth network, namely Classifier-1 from \citetalias{shu22}, was primarily targeting high-redshift strong-lenses and trained on negative examples drawn from a parent sample with red ($g-r$) and ($g-i$) colors. All five networks were trained and validated on images in $gri$ bands, such that we require only the availability of $gri$ bands in HSC, while the image in $z$ or $y$ can be missing.

\FloatBarrier
\section{Cleaning the output candidate list}
\label{sec:cleaning}

Before conducting a visual inspection (see Sect.~\ref{sec:visualinspection}) to identify the high-quality strong lens candidates, the catalog of 22,393 candidates was post-processed to remove obvious artifacts and non-lenses. Since prior to this work no larger samples of HSC images were flagged as images with artifacts or crowded fields, we could not include a larger fraction of them in the negative training set. This would have helped to lower the FP rate. However, it would have lowered the fraction of other types (such as LRGs, spirals, or ring galaxies; see \citetalias{canameras24}) in the negative set that more closely resemble the lens images. For this post-processing, we applied mainly two criteria, as is detailed below. An overview table of the stages and resulting sample sizes is given in Table~\ref{tab:summary}. While both techniques could have also been applied to the $\sim$110 million parent sample and only the cleaned sample would have been classified by the network committee, it is significantly more efficient to first rank the full sample, and then further clean only the top candidates. This is mostly due to the longer runtime of the cleaning scripts than a network evaluation and because the second approach (see Sect.~\ref{sec:cleaning:model}) requires in addition to the $gri$ bands the $z$ band observations which can then be downloaded only for the top ranked candidates rather than the full $\sim$110 million parent sample.

\begin{table}[t!]
    \caption{Summary of the different selection stages with corresponding sample size, including rediscoveries.}
    \begin{center}
    \begin{tabular}{lcc}\hline \hline
Description   & Sample size & Section \\   \hline \noalign{\smallskip}
Parent sample (in million) & $\sim$110 & \ref{sec:network}\\
Network candidates & 22,393 & \ref{sec:network} \\ \hline
After SExtractor cleaning & 19,820 & \ref{sec:cleaning:sextractor}\\
After pixel level cuts & 18,712 & \ref{sec:cleaning:sextractor}\\
After CNN lens model cleaning & 15,919 & \ref{sec:cleaning:model}\\
After excluding duplicates ($<2\arcsec$) & 14,448 & \ref{sec:visualinspection}\\
After excluding previous inspected systems & 11,773 & \ref{sec:visualinspection}\\ \hline
Inspected in binary classification round & 11,874 & \ref{sec:visualinspection:binary} \\ 
Inspected by first team & 3,408 & \ref{sec:visualinspection:multi}\\
Recentered after first team's inspection & 160 & \ref{sec:visualinspection:multi} \\
Inspected by second team & 686 & \ref{sec:visualinspection:multi}\\ \hline 
Total inspected systems listed in Table~\ref{tab:newcand} & 14,152 & \ref{sec:visualinspection:final}\\  
New published grade A (B) lens candidates & 9 (83) & \ref{sec:visualinspection:final}\\ 
New inspected grade A (B) lens candidates & 42 (187) & \ref{sec:visualinspection:final}\\ 
All grade A (B) lens candidates & \gradeA (\gradeB) & \ref{sec:visualinspection:final}\\
Grade A (B) lens candidates with OD$_\text{vis}$ & 22 (66) &\ref{sec:visualinspection:discussion}\\
Grade A (B) lens candidates with OD$_\text{z}$ & 18 (70) & \ref{sec:visualinspection:discussion}\\ 
New inspected grade A with good model & 28 (67\%) &\ref{sec:visualinspection:discussion} \\ 
New inspected grade B with good model & 164 (88\%) &\ref{sec:visualinspection:discussion} \\ 
Inspected grade A or B with good model & 192 (84\%) &\ref{sec:visualinspection:discussion} \\ \hline
\end{tabular}
    \end{center}
    \label{tab:summary}
\end{table}

\subsection{Cleaning using SExtractor and HSC pixel level flags}
\label{sec:cleaning:sextractor}

\begin{figure}[t!]
\centering
\includegraphics[width=.5\textwidth]{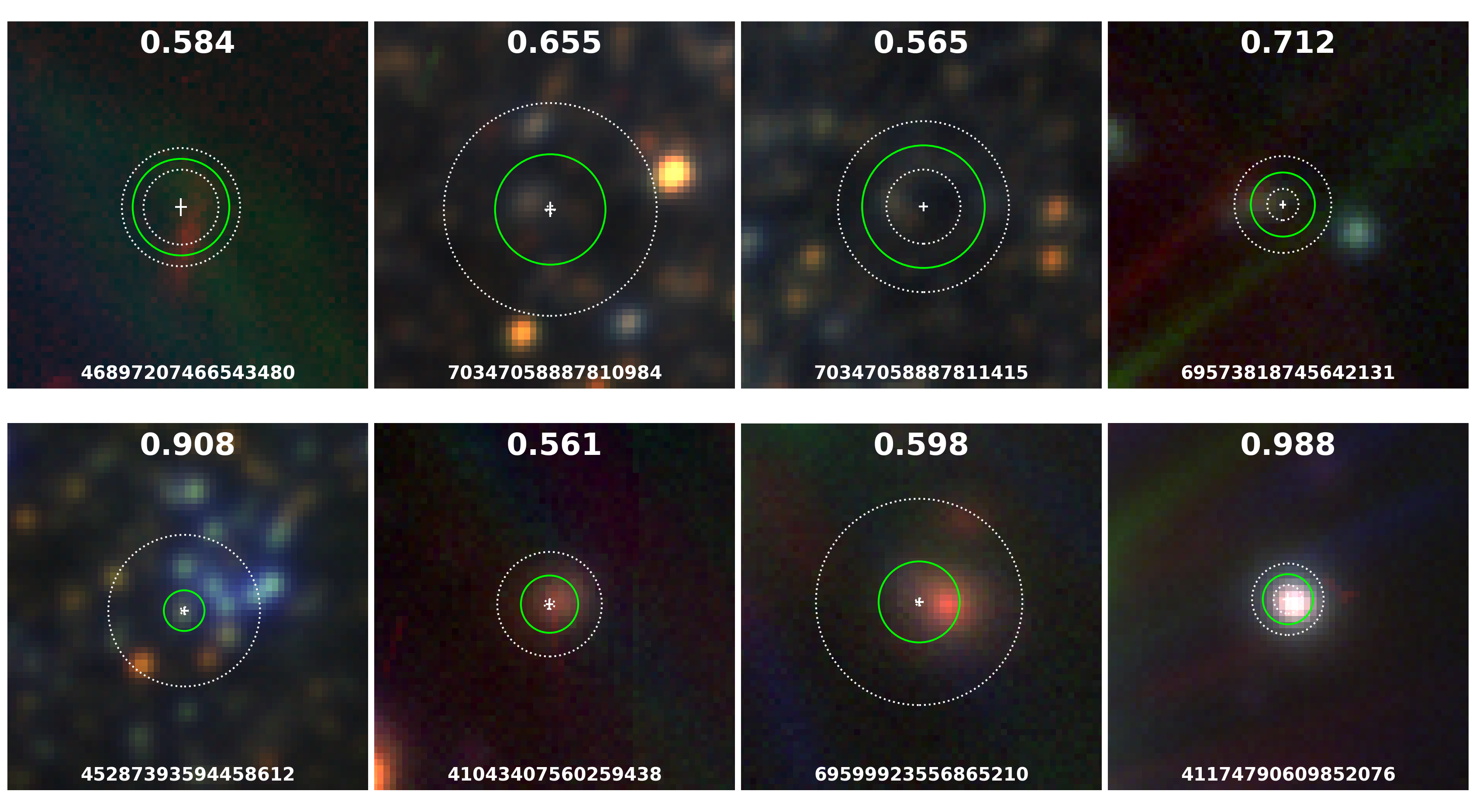}
\caption{Examples of false-positive candidates from the network committee that were cleaned with post-processing scripts. A major fraction of contaminants correspond to crowded fields and/or cutouts with nonzero background emission in $g$, $r$, and/or $i$-band, which are rejected by SExtractor and HSC pixel-level flags (top row) or by the modeling network from \citetalias{schuldt23a} (bottom row). We show the average score of the network committee and the HSC ID in the top and bottom, respectively, of each panel, which is $\sim 10\arcsec$ on a side. We further show the predicted lens center as a cross with a length corresponding to the predicted uncertainty, and the Einstein radius as a solid green circle with predicted uncertainty bounds as dotted white lines. Some of the lower limits on the Einstein radii are near zero and are thus not visible.}
\label{fig:ex_cleaning}
\end{figure}

A substantial fraction of contaminants correspond to cutouts with residual background emission in one or all $gri$ bands (see some examples in the top row of Fig. \ref{fig:ex_cleaning}). These cutouts were identified using source masking and estimates of the sky background with SExtractor \citep{bertin96}, leaving 19,820 candidates without loss of any test lens. This step was then refined with information from the pixel-level flags inferred by the HSC pipeline. We searched for optimal cuts to remove contaminants using the flags in $gri$ bands over the 72\,$\times$\,72 pixel images. First, we discarded cutouts that have $>$90\% pixels with flags $\geq$512 and $<$1024, corresponding to pixels located near a bright object. This securely excluded cutouts affected by bright neighboring stars within a few tens of arcseconds. Secondly, we removed cutouts with $>$99\% pixels flagged as “detected pixels”, which are more likely to correspond to crowded fields such as stellar clusters, or star-forming clumps within extended disks, than strong gravitational lenses. These two criteria decreased the sample to 18,712 candidates.

\subsection{Cleaning using modeling CNN}
\label{sec:cleaning:model}

In parallel, we applied the ResNet from \citetalias{schuldt23a} to all 22,393 network candidates. This modeling network is trained on realistic mock images created in a similar way to the ones used for training the committee network, and obtained a great performance in measuring the Einstein radii of lens systems when compared to \gleeautopy \citep{schuldt23b}, a code that relies on the well-tested lens modeling software \GLEE \citep{suyu10a_GLEE, suyu12b_GLEE}. The modeling network predicts the lens mass center, $x$ and $y$, the lens mass ellipticity, $e_\text{x}$ and $e_\text{y}$, and Einstein radius, $\theta_\text{E}$, of a SIE profile as well as an external shear ($\gamma_1$ and $\gamma_2$), and the corresponding 1$\sigma$ uncertainties. 

\begin{figure*}[t!]
\centering
\includegraphics[trim={120 18 95 50},clip, width=\textwidth]{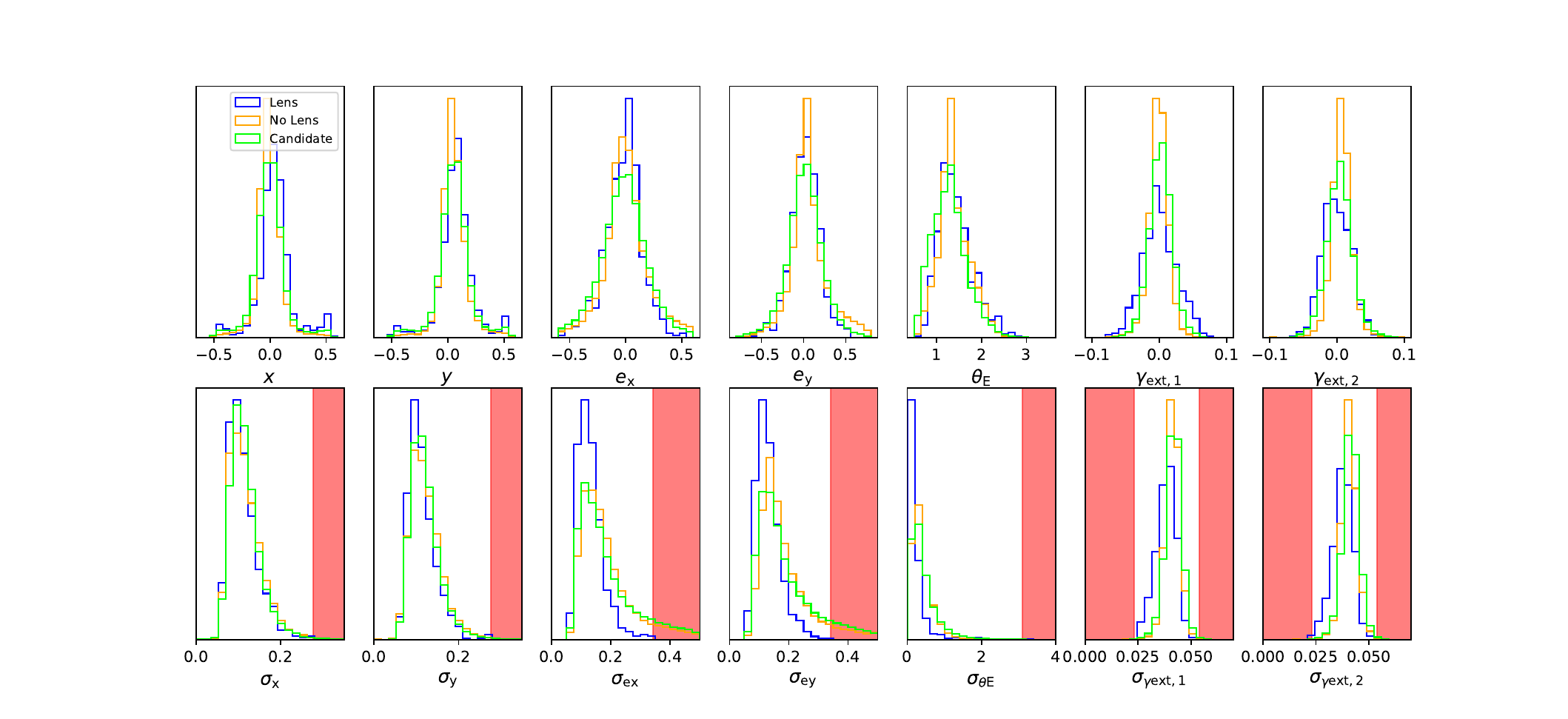}
\caption{Normalized histograms of the mass model parameter predictions using the modeling network from \citetalias{schuldt23a}.  The top row shows the predicted mass model parameter values, and the bottom row shows the corresponding 1$\sigma$ uncertainties. We show the 546 grade A and B lens candidates from \citetalias{canameras21b} and \citetalias{schuldt25a} (blue), the visually rejected systems as non-lenses for comparison (orange), and those of the network candidates from our committee network of this work (green). The parameter ranges used to reject non-lenses are marked in red and defined by the highest and/or  lowest uncertainty of the above-mentioned grade A and B lens sample (the small overlaps between the red regions and nonzero blue bins are merely due to the finite bin widths of the histograms).}
\label{fig:modelcuts}
\end{figure*}

Since the modeling network is trained on solely lens images (i.e., not in combination with lens classification as in \citet{andika25}), it is forced to provide reasonable model parameters (e.g., $\theta_\text{E} \in $[0\farcs5, $5\arcsec]$) even if the given image is clearly not a lens. However, for such a given non-lens image, the modeling network is uncertain and the predicted errors can be significantly higher. Consequently, we can use this fact to reject false-positives from the classification network. To be conservative, we define the cuts here such that we do not exclude any lens candidate from our test set, as is shown in Fig.~\ref{fig:modelcuts}, which results in an additional rejection of 2,794 systems. We show some examples of rejected candidates in the bottom row of Fig.~\ref{fig:ex_cleaning}. 

In sum, we rejected around 29\% of the network candidates as false positives by keeping the same TPR on our test sample. This cleaning was performed through automated and fast scripts, and thus is also scalable for sample sizes expected from ongoing and upcoming wide-field imaging surveys that are two orders of magnitude higher. While excluding $\sim 1/3$ of the contaminants by keeping the same TPR is already a good improvement in the purity, we note that for significantly higher samples more stringent cuts can be applied to reject more contaminants, with the downside of possibly excluding some lenses. 

\section{Visual inspection}
\label{sec:visualinspection}

Before the visual inspection of the remaining network candidates from Sect.~\ref{sec:cleaning}, we excluded duplicates among the network candidates. Although CNNs are known to be translation-invariant, we only excluded duplicates within two pixels when creating the 110 million parent sample to be very conservative. This ensured that we got network scores for all individual galaxies in the parent sample, and that we preserved cutouts centered on the actual lens galaxies. However, for the visual inspection, we removed duplicate cutouts within 2\arcsec. In order to preserve lines of sight with the most extended and most likely deflector galaxy in the center, we rank ordered the catalog entries by decreasing $i$-Kron radius and removed duplicates of lower rank. Excluding these duplicates reduces the human inspection time further, while the moderate cut of 2\arcsec ensures that we will not shift any multiple image out of the inspected cutout, which would risk us missing the identification. 

Furthermore, we excluded additional image stamps previously inspected as part of our HOLISMOKES strong lens searches in earlier HSC data releases (\citetalias{canameras21b}; \citetalias{shu22}; \citet{schuldt25a}, hereafter \citetalias{schuldt25a}) to lower the need in human resources, as we grade the systems in a similar way and would therefore have similar resulting grades. We only kept 50 lenses (high-quality grade A or B lens candidates, which we simply refer to as ``lenses'' hereafter) and 50 non-lenses from \citetalias{canameras21b} and \citetalias{schuldt25a} relying both on the same network, as well as 50 lenses and 50 non-lenses from \citetalias{shu22} for comparison\footnote{We use here a lower limit of the threshold score $p=0.53$ to include also a few systems that would be usually excluded with the cut at $p=0.55$.}. This resulted in a catalog of 11,874 candidates, including the 200 systems mentioned above, for visual inspection. 

The visual inspection closely followed the procedure proposed by \citetalias{schuldt25a} and was conducted jointly with candidates presented by \citet{andika25}. In short, we (1) carried out a calibration round using 200 systems inspected by all eight inspectors\footnote{The visual inspectors are in alphabetical order by last name: I.~T.~A., S.~B., R.~C., C.~G., A.~M., S.~S., S.~H.~S., and S.~T.} since we introduced several new features in the grading tool, as is detailed below. The actual grading then comprised (2) a binary classification (Sect.~\ref{sec:visualinspection:binary}), a (3) four-grade inspection of four individuals, and (4) another four-grade inspection from the remaining inspectors, which additionally characterized the predicted mass model and the candidate environment (see Sect.~\ref{sec:visualinspection:multi}). Finally, we obtained the average grade, $G$, from the eight individual grades collected for the interesting candidates (reported in Table~\ref{tab:newcand}). As previously, we use a three-panel image to show two different scalings of the $gri$ bands, and, in contrast to previous inspections, an image of the $riz$ bands to ease the identification of lensed quasars. While these image stamps have a size of $10\arcsec \times 10\arcsec$, we additionally show $80\arcsec \times 80\arcsec $ stamps in the same filters and scaling as in the smaller cutout. 

\subsection{Binary inspection}
\label{sec:visualinspection:binary}

We then split the catalog among ourselves, such that each system was inspected by two graders in a binary classification to rapidly exclude the majority of obvious non-lenses. As it was mentioned above, the catalog contains 200 systems from our previous searches, with 100 lenses and 100 non-lenses. Of the 100 lenses (or lens candidates), we recovered 98/100 with a previous grade of A or B and only missed two systems, namely systems HSC J222002+060506\footnote{Listed as HSC J2220+0605 in \citetalias{canameras21b}.} and HSC J090929+010030 which previously obtained both an average grade of 1.6, while the lower limit for grade B is 1.5. Of the 100 non-lenses, we forwarded 59/100, which, however, is understandable as several of these galaxies previously obtained an average grade slightly below 1.5 and we aim to now forwarded all systems above a grade of 1 in the final grading scheme. Interestingly, we also forwarded 12/100 that previously had an average grade of 0. Since this first stage is graded conservatively to rule out obvious non-lenses and there is subsequently another round of inspection for refinement, this result for the 200 systems is overall very good. 

\subsection{Multi-class inspection}
\label{sec:visualinspection:multi}

In the next round, four individuals inspected the remaining 3,408 systems from the binary classification. At this stage, each inspector provided one of four possible grades ($0$ corresponding to ``no lens'', $1$ to ``possible lens'', $2$ to ``probable lens'', and $3$ to ``definite lens'') and voted in case the lens was significantly offset from the cutout center. In total, 309 systems obtained the ``off-center" flag by at least one person, such that the lens center was subsequently corrected. A re-centering at this stage was crucial for the subsequent round of grading by the other four remaining graders, as we also showed in the central image the lens center and Einstein radius predicted by the ResNet from \citetalias{schuldt23a} (see bottom row of Fig.~\ref{fig:ex_cleaning} for examples).

\begin{figure}[htb!]
\centering
\includegraphics[width=.5\textwidth]{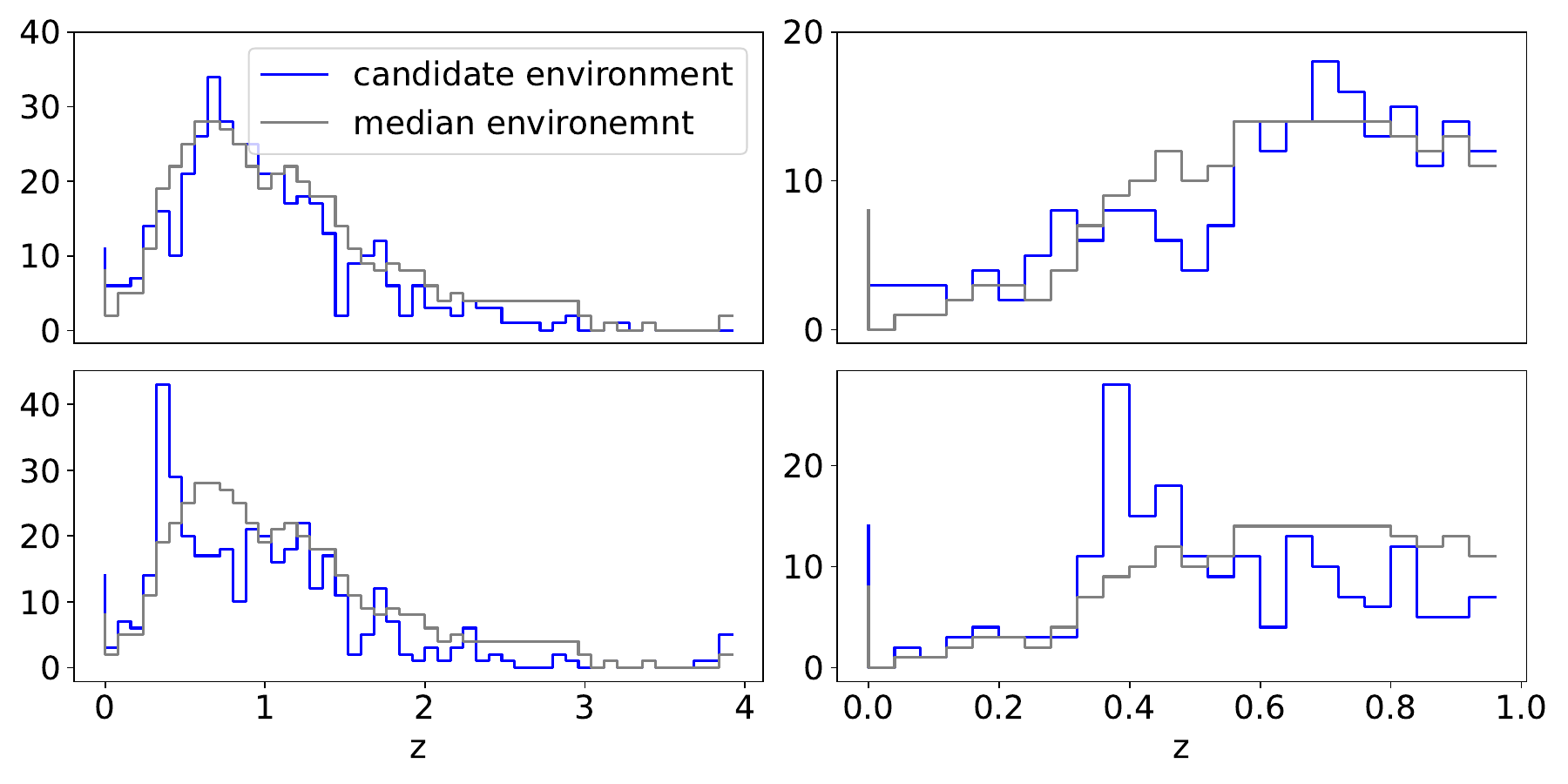}
\caption{Histograms of the photometric redshifts within a box of 200\arcsec on a side around two different lens candidates (top and bottom row) as examples. Such histograms, one in the range up to a redshift of 4 (left) and one up to a redshift of 1 (right), were shown during the visual inspection to ease the environment classification.  The lens system in the top row has an environment similar to the median, whereas the system in the bottom row shows an overdensity at $z\sim0.4$.  }
\label{fig:ex_hist}
\end{figure}

Based on the average grade, $G$, obtained from these four grades, we forwarded all systems with either $G>1$ or $G\leq 1$ but a standard deviation of the four grades above 0.75. Including those with a lower grade but a high discrepancy increases the forwarded sample significantly, but minimizes the possibility of missing any good lens candidate. This results in 686 candidates, including 160 systems that got shifted, which were inspected by the remaining four graders.  In addition to the grading and ``off-center" flag from the previous round, these four individuals were tasked with also indicating if the predicted lens center or Einstein radius is significantly mis-predicted. Furthermore, for each candidate, two histograms of photometric redshifts from the lens candidate environment, one up to $z=1$ and the other up to $z=4$, were shown in order to obtain a classification of the environment as well. As examples, the histograms of one system in an overdensity (bottom row) and one in the field (top row) are shown in Fig.~\ref{fig:ex_hist}. This allows us to assess if the predicted mass model, which we provide in Table~\ref{tab:newcand} (see also Table~\ref{tab:legend} for explanations), is reliable and to characterize the systems' environment, while only very slightly increasing the visual inspection time.

\begin{table*}[t!]
    \caption{Lens candidates with network committee scores of $p\geq 0.55$ that were visually inspected (excluding duplicates). The full table is available in electronic form at the SuGOHI and HOLISMOKES webpages, and CDS, and each column content is explained in Table~\ref{tab:legend}.}
    \begin{center}
    \begin{tabular}{c|cc|cccc|ccc|c|l}
Name   &RA [deg]  &Dec [deg]  & $p$ & $G$ & $\sigma_\text{G}$ & $N_\text{graders}$ & $\theta_\text{E}$ [\arcsec] & $\sigma_{\theta_\text{E}}$ & Model & $z$ & References\\
(1)& (2) & (3) & (4) & (5) & (6) & (7) & (24) & (25) & (30) & (31) & (37) \\ \hline \hline \noalign{\smallskip}
 HSCJ1004$-$0031 & $ 151.21543 $ & $ -0.52911 $ & $ 0.64 $ & $ 3.00 $ & $ 0.00 $ & 8 & 1.44 & 0.15 & Y & 1.05 & A23\\
 HSCJ2305$-$0002 & $ 346.34026 $ & $ -0.03658 $ & $ 0.60 $ & $ 3.00 $ & $ 0.00 $ & 8 & 1.32 & 0.12 & Y & 0.49 & W18 C21 \\
 HSCJ2242$+$0011 & $ 340.58995 $ & $  0.19573 $ & $ 0.71 $ & $ 3.00 $ & $ 0.00 $ & 8 & 1.51 & 0.11 & Y & 0.38 & S18 R22 \\
 HSCJ1236$-$0035 & $ 189.15056 $ & $ -0.59418 $ & $ 0.86 $ & $ 3.00 $ & $ 0.00 $ & 8 & 0.90 & 0.06 & Y & 0.49 & S22 \\
 HSCJ0102$+$0159 & $  15.65957 $ & $  1.98211 $ & $ 0.61 $ & $ 3.00 $ & $ 0.00 $ & 8 & 1.38 & 0.09 & N & 0.95 & J19 C21\\
 \vdots & \vdots &  \vdots &  \vdots &  \vdots &  \vdots &  \vdots &  \vdots &  \vdots &  \vdots &  \vdots &  \vdots \\ \hline
    \end{tabular}
    \end{center}
    \label{tab:newcand}
\end{table*}

\subsection{Comparison to previous inspected candidates}
\label{sec:visualinspection:comparison}

As was mentioned above, we included 200 systems that we already inspected in previous works. From the 100 candidates that we previously excluded (i.e., that obtained an average grade of $G<1.5$), only one system has now obtained an average grade above 1.5 and is consequently listed as a grade B candidate. From the sample of previously classified grade A and B lenses, we notice a tendency toward lower grades among most graders. While there were some changes compared to previous inspections, such as moving from PDR2 to PDR3 and showing two $gri$ color images and one $riz$ color image instead of three $gri$ color images, we checked these aspects by directly comparing some inspected images from \citetalias{schuldt25a} and this work (see example images in Fig.~\ref{fig:PDR3vsPDR2}), and do not see this as a major source of bias. We also note that the applied scalings depend on the pixel values of the given cutout, such that a shifted image appears slightly different. Additionally, in \citetalias{shu22} we focused on high-$z$ lenses and applied slightly different scaling functions, but find this to be not a major reason. Furthermore, we speculate that this tendency may come from the relatively pure sample (i.e., a higher proportion of lenses) that we had after the binary inspection, but note that we also had in previous works a binary inspection cleaning carried out by a single inspector, resulting in a comparably pure sample. In contrast, we noticed that the 80\arcsec$\times$80\arcsec cutouts help in specific cases to reject non-lenses as the overall environment is visible. Finally, we remark that it is known \citep{rojas23, schuldt25a} that every grader has a different expectation and also the grades among a single inspector vary. Consequently, it might be simply that our expectations of an object qualifying as a lens candidate have slightly increased over time, possibly because of the increasing sample of lenses and the higher availability of high-resolution images (e.g., Euclid, see \citet{acevedobarroso25, nagam25, walmsley25}).

\subsection{Final lens compilation}
\label{sec:visualinspection:final}

\begin{figure*}[thb!]
\centering
\includegraphics[width=\textwidth]{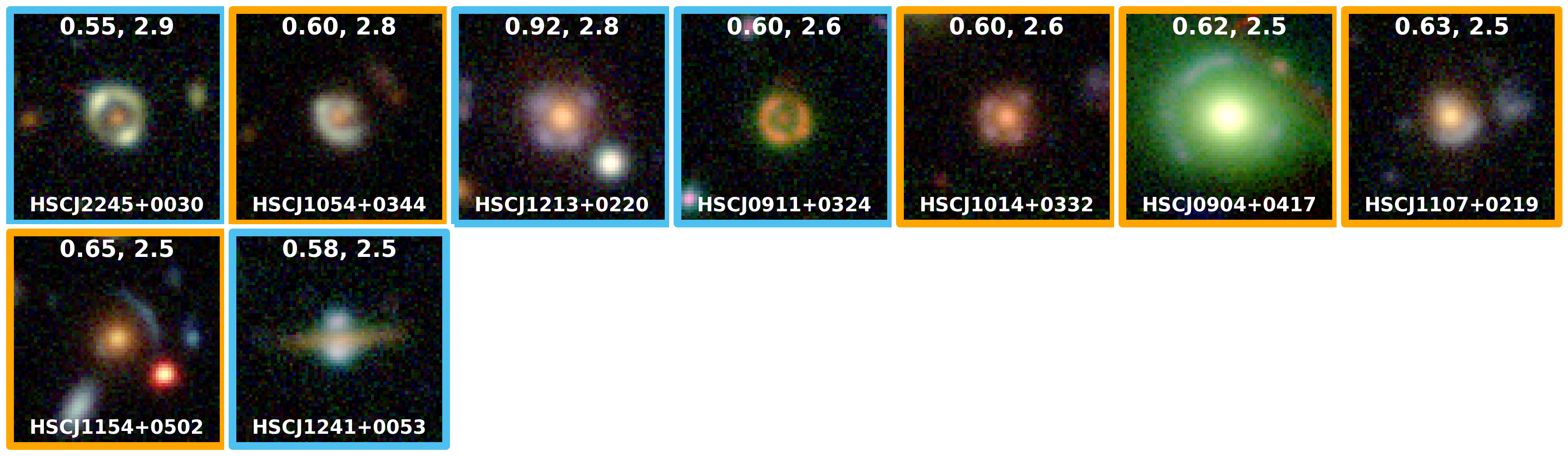}
\caption{Color-image stamps ($12\arcsec \times 12\arcsec$; north is up and east is left) of identified grade-A lens candidates using HSC PDR3 $gri$ multiband imaging data that are detected for the first time. At the top of each panel, we list the scores of the network committee, $p$, and the average visual inspection grade, $G$, of eight graders, where $\geq 2.5$ corresponds to grade A. At the bottom, we list the candidate name. We further distinguish between candidates discovered jointly with \citet{andika25} using VariLens, appearing with blue frames, and fully new identifications with orange frames. 
All systems with their coordinates, and further details such as the lens environment and Einstein radii, are listed in Table~\ref{tab:newcand}.}
\label{fig:gradeA}
\end{figure*}

\begin{figure*}[thb!]
\centering
\includegraphics[width=0.91\textwidth]{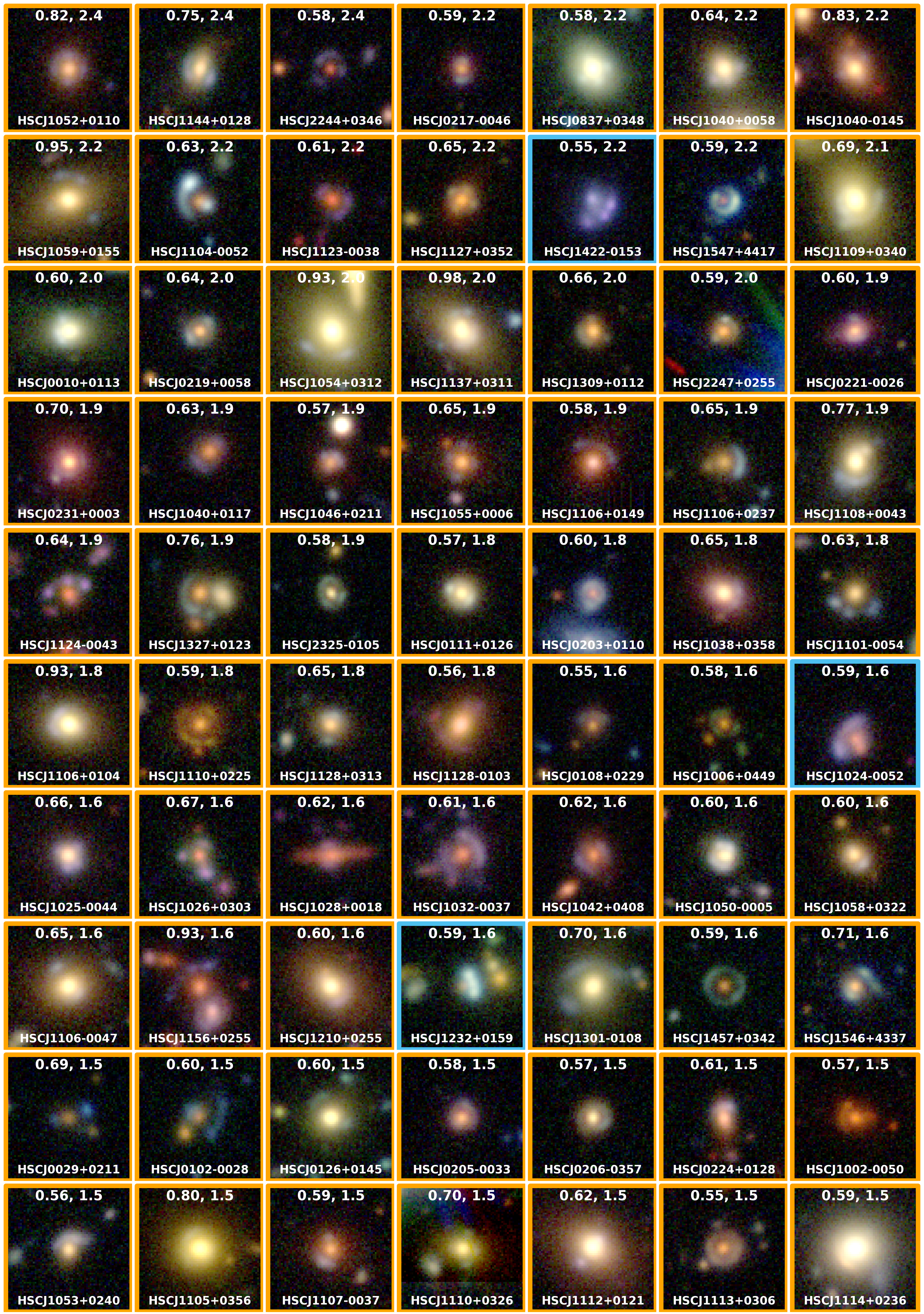}
\caption{Color-image stamps of newly identified grade-B lens candidates, in the same format as Fig.~\ref{fig:gradeA}}
\label{fig:gradeB}
\end{figure*}

Despite the tendency toward lower grades, we adopt our traditional cuts to define grade A ($3\geq G \geq 2.5$) and grade B ($2.5 > G \geq 1.5$), and release in Table~\ref{tab:newcand} the full catalog of visually inspected systems. This catalog includes jointly inspected systems that obtained a committee score of $p\geq 0.53$ and were also detected by VariLens \citep[][noted in column 12 of the catalog]{andika25}, but we exclude any duplicates within $<2$\arcsec. Consequently, using the average grades, $G$, of all eight inspectors to obtain a stable average \citep[see also][]{rojas23, schuldt25a}, we found 42 grade A  and 187 grade B lens candidates, which are listed in Table~\ref{tab:newcand}. Furthermore, the newly discovered grade A candidates are shown in Fig.~\ref{fig:gradeA}, while the grade B systems are shown in Fig.~\ref{fig:gradeB}. We further provide in this table all grade A and B lens candidates from \citetalias{canameras21b} (marked in column 15, see Table~\ref{tab:legend}), \citetalias{shu22} (marked in column 14), and \citetalias{schuldt25a} (marked in column 13) that we detected with our network committee as candidates but excluded from visual inspection (see also Sect.~\ref{sec:visualinspection:multi}). In sum, we found with our network committee 95 grade A and 503 grade B lenses. From these systems, \rediscoveries systems (included not regraded systems) were already known, while we show in Figs.~\ref{fig:gradeA_lit} (grade A) and \ref{fig:gradeB_lit} (grade B) those that we regraded in our inspection (either because they are in the small comparison sample, see Sect.~\ref{sec:visualinspection:comparison}, or because they were discovered by other work and not by \citetalias{canameras21b}, \citetalias{shu22}, or \citetalias{schuldt25a}). This high recovery rate is expected, given the enormous lens search projects that already exploited HSC data and in particular our previous searches using individual networks that entered into our network committee now. However, we remind the reader that the new lens identification is only one aspect of this work. 

\begin{figure*}[thb!]
\centering
\includegraphics[width=\textwidth]{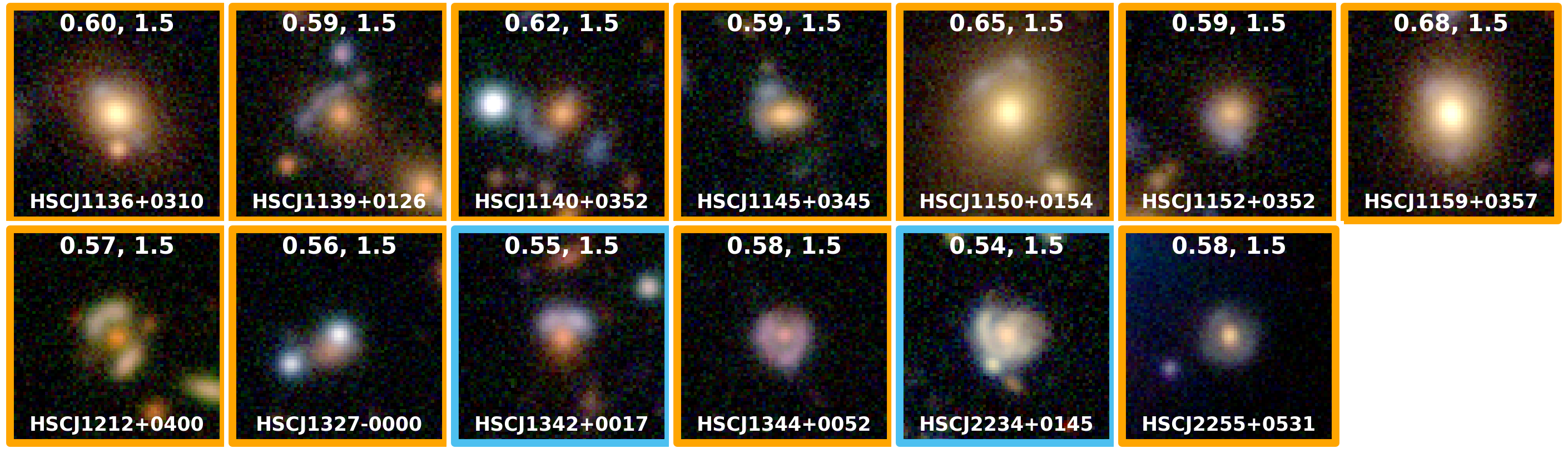}
\caption*{Fig.~\ref{fig:gradeB} (continued): Color-image stamps of newly identified grade-B lens candidates, in the same format as Fig.~\ref{fig:gradeA}}
\end{figure*}

\subsection{Discussion on the lens properties}
\label{sec:visualinspection:discussion}

We further classified the lens environment using photo-$z$ histograms during the inspection by the second team (see Fig.~\ref{fig:ex_hist} for examples), leading to an identification of 147 candidates in an overdensity (corresponding to at least two votes), of which 22 and 66 are ranked as grade A and B candidates, respectively. With the automated photo-$z$ procedure to identify overdensities presented by \citetalias{schuldt25a}, we identify 569 systems among the whole inspected sample, of which 18 and 70 are classified as grade A or B lens candidates, respectively (see also Tab.~\ref{tab:summary}). While the automated procedure can be easily applied to large samples, the visual classification requires in particular human time that is highly limited. On the other hand, we visually identified several systems as being in an overdensity that were missed by the automated procedure due to missing $z$ or $y$ band observations required for the photo-$z$ codes. In total, we find a good overlap between the two complementary procedures.

As was mentioned earlier, we further showed the network-predicted lens center and Einstein radius during final inspection. Since we found in the calibration round that the model network from \citet{schuldt23a} has difficulties in correctly predicting the model if the lens is not well centered, we mitigated this through a re-centering between the last two inspection stages. We overlay the predicted lens center and Einstein radius on the cutouts of our lens candidates in Appendix~\ref{sec:appendixB}. For the reported lens center coordinates in Table~\ref{tab:newcand}, we only found 9\% of the lens candidates to have a relatively poor predicted lens center or Einstein radius, which might be because of a group- or cluster-scale lens not being suited for the modeling network or a remaining offset of the lens. We provide the predicted values of the seven mass model parameters with their corresponding 1 $\sigma$ uncertainties in Table~\ref{tab:newcand} (columns 16 to 29, see also Table~\ref{tab:legend}) as well as a flag (column 30) if the lens center and Einstein radius match the system reasonably (defined as two or more votes). This demonstrates once more the power of the modeling network and shows that the result can be used for further analysis. 

While we applied in Sect.~\ref{sec:cleaning:model} relatively weak cuts on the model parameter uncertainties (see Fig.~\ref{fig:modelcuts}), we note that more stringent cuts are possible when the sample sizes become significantly larger. This would lead to an even higher fraction of systems that can be easily ruled out, while keeping most of the lens candidates.


\section{Summary and conclusion}
\label{sec:conclusion}

We have carried out a systematic search for strong gravitational lenses using the $gri$ bands of the HSC Wide layer observations from the third public data release. For this, we tested the combination of different networks by averaging or multiplying their individual network scores. The best performance, 75\% completeness on real lenses at a FPR of $\sim 0.01\%$, was obtained by averaging the scores from five different networks. 

While it would easily have been possible to visually inspect the resulting sample size, this will change with the next generation of wide-field imaging surveys, such that further cleaning is unavoidable. We tested two approaches, first using SExtractor and the HSC pixel-level flags to reject images with mostly nonzero background or image artifacts, and second, the ResNet modeling network from \citet{schuldt23a} as it predicts higher uncertainties for non-lenses. This lowered the contaminants by 29\%, while not excluding any lens candidate from our previous identifications. Such post-processing scripts are expected to play a crucial role when two orders of magnitude more images need to be classified. 

Thanks to a visual inspection of the cleaned network candidate list, we identified \gradeA grade A and \gradeB grade B candidates, including \rediscoveries previously known systems. We provide in Table~\ref{tab:newcand} their coordinates together with their SIE and external shear parameters obtained from the ResNet presented by \citet{schuldt23a}. During the visual inspection, we also showed for the first time the predicted lens center and Einstein radius, and characterized the reliability. We found that only $\sim9$ \% of the candidates obtained a lens center or Einstein radius that is not well predicted, once the system is well centered. Moreover, we characterized their environment with two complementary approaches, either through a visual inspection of histograms showing the photo-$z$ values of their surroundings, or the selection cuts elaborated by \citetalias{schuldt25a}. In both cases, we find 88 grade A or B systems to be in a significantly overdense environment. In Table~\ref{tab:newcand}, which includes all visually inspected candidates with our obtained average grades and their characteristics as well as their previous discoveries, we are releasing one of the most complete catalogs of strong lensing systems so far.

Particularly in preparation for the ongoing and upcoming wide-field imaging surveys, the removal of false-positives of network classifiers will be a challenge, so the approaches presented here are expected to play a crucial role. Developing and testing fast and autonomous techniques to characterize and analyze the new systems, such as the ones tested here as well, will be a necessity.

\section*{Data Availability}
Table~\ref{tab:newcand} is available at the CDS via anonymous ftp to
\url{cdsarc.cds.unistra.fr} (130.79.128.5) or via \url{http://cdsweb.u-strasbg.fr/cgi-bin/qcat?J/A+A/}. It was further released through the HOLISMOKES collaboration webpage at \url{www.holismokes.org} and accordingly incorporated in the SuGOHI data base accessible through \url{https://www-utap.phys.s.u-tokyo.ac.jp/~oguri/sugohi/}.

\begin{acknowledgements}
We thank the anonymous referee for comments that helped to improve the clarity of the paper.
SS has received funding from the European Union’s Horizon 2022 research and innovation programme under the Marie Skłodowska-Curie grant agreement No 101105167 — FASTIDIoUS. We acknowledge financial support through grants PRIN-MIUR 2017WSCC32 and 2020SKSTHZ. 
This project has received funding from the European Research Council (ERC) under the European Union's Horizon 2020 research and innovation programme (LENSNOVA: grant agreement No 771776). This research is supported in part by the Excellence Cluster ORIGINS which is funded by the Deutsche Forschungsgemeinschaft (DFG, German Research Foundation) under Germany's Excellence Strategy -- EXC-2094 -- 390783311. 
This work uses the following software packages:
\href{https://github.com/astropy/astropy}{\texttt{Astropy}}
\citep{astropy1, astropy2},
\href{https://github.com/matplotlib/matplotlib}{\texttt{matplotlib}}
\citep{matplotlib},
\href{https://github.com/numpy/numpy}{\texttt{NumPy}}
\citep{numpy1, numpy2},
\href{https://www.python.org/}{\texttt{Python}}
\citep{python},
\href{https://github.com/scipy/scipy}{\texttt{Scipy}}
\citep{scipy}.
\end{acknowledgements}

\bibliographystyle{aa}
\bibliography{main}

\begin{thebibliography}{101}
\expandafter\ifx\csname natexlab\endcsname\relax\def\natexlab#1{#1}\fi

\bibitem[{{Acebron} {et~al.}(2022){Acebron}, {Grillo}, {Bergamini}, {Caminha},
  {Tozzi}, {Mercurio}, {Rosati}, {Brammer}, {Meneghetti}, {Nonino}, \&
  {Vanzella}}]{acebron22_SDSS2222}
{Acebron}, A., {Grillo}, C., {Bergamini}, P., {et~al.} 2022, \aap, 668, A142

\bibitem[{{Acebron} {et~al.}(2024){Acebron}, {Grillo}, {Suyu}, {Angora},
  {Bergamini}, {Caminha}, {Ertl}, {Mercurio}, {Nonino}, {Rosati}, {Wang},
  {Bolamperti}, {Meneghetti}, {Schuldt}, \&
  {Vanzella}}]{acebron24_extendedmodel}
{Acebron}, A., {Grillo}, C., {Suyu}, S.~H., {et~al.} 2024, \apj, 976, 110

\bibitem[{{Acevedo Barroso} {et~al.}(2025){Acevedo Barroso}, {O'Riordan},
  {Cl{\'e}ment}, {Tortora}, {Collett}, {Courbin}, {Gavazzi}, {Metcalf},
  {Busillo}, {Andika}, {Cabanac}, {Courtois}, {Crook-Mansour}, {Delchambre},
  {Despali}, {Ecker}, {Franco}, {Holloway}, {Jackson}, {Jahnke}, {Mahler},
  {Marchetti}, {Matavulj}, {Melo}, {Meneghetti}, {Moustakas}, {M{\"u}ller},
  {Nucita}, {Paulino-Afonso}, {Pearson}, {Rojas}, {Scarlata}, {Schuldt},
  {Serjeant}, {Sluse}, {Suyu}, {Vaccari}, {Verma}, {Vernardos}, {Walmsley},
  {Bouy}, {Walth}, {Powell}, {Bolzonella}, {Cuillandre}, {Kluge}, {Saifollahi},
  {Schirmer}, {Stone}, {Acebron}, {Bazzanini}, {D{\'\i}az-S{\'a}nchez}, {Hogg},
  {Koopmans}, {Kruk}, {Leuzzi}, {Manj{\'o}n-Garc{\'\i}a}, {Mannucci}, {Nagam},
  {Pearce-Casey}, {Scharr{\'e}}, {Wilde}, {Altieri}, {Amara}, {Andreon},
  {Auricchio}, {Baccigalupi}, {Baldi}, {Balestra}, {Bardelli}, {Basset},
  {Battaglia}, {Bender}, {Bonino}, {Branchini}, {Brescia}, {Brinchmann},
  {Caillat}, {Camera}, {Candini}, {Capobianco}, {Carbone}, {Carretero},
  {Casas}, {Castellano}, {Castignani}, {Cavuoti}, {Cimatti}, {Colodro-Conde},
  {Congedo}, {Conselice}, {Conversi}, {Copin}, {Corcione}, {Cropper}, {Da
  Silva}, {Degaudenzi}, {De Lucia}, {Dinis}, {Dubath}, {Dupac}, {Dusini},
  {Farina}, {Farrens}, {Ferriol}, {Frailis}, {Franceschi}, {Galeotta},
  {Garilli}, {George}, {Gillard}, {Gillis}, {Giocoli}, {G{\'o}mez-Alvarez},
  {Grazian}, {Grupp}, {Guzzo}, {Haugan}, {Hoekstra}, {Holmes}, {Hook},
  {Hormuth}, {Hornstrup}, {Jhabvala}, {Joachimi}, {Keih{\"a}nen}, {Kermiche},
  {Kiessling}, {Kubik}, {Kunz}, {Kurki-Suonio}, {Le Mignant}, {Ligori},
  {Lilje}, {Lindholm}, {Lloro}, {Mainetti}, {Maiorano}, {Mansutti}, {Marcin},
  {Marggraf}, {Martinelli}, {Martinet}, {Marulli}, {Massey}, {Medinaceli},
  {Melchior}, {Mellier}, {Merlin}, {Meylan}, {Moresco}, {Moscardini}, {Munari},
  {Nakajima}, {Neissner}, {Nichol}, {Niemi}, {Nightingale}, {Padilla},
  {Paltani}, {Pasian}, {Pedersen}, {Percival}, {Pettorino}, {Pires}, {Polenta},
  {Poncet}, {Popa}, {Pozzetti}, {Raison}, {Rebolo}, {Renzi}, {Rhodes},
  {Riccio}, {Romelli}, {Roncarelli}, {Rossetti}, {Saglia}, {Sakr},
  {S{\'a}nchez}, {Sapone}, {Schneider}, {Schrabback}, {Secroun}, {Seidel},
  {Serrano}, {Sirignano}, {Sirri}, {Skottfelt}, {Stanco}, {Steinwagner},
  {Tallada-Cresp{\'\i}}, {Tavagnacco}, {Taylor}, {Tereno}, {Toledo-Moreo},
  {Torradeflot}, {Tutusaus}, {Valentijn}, \& {Valenziano}}]{acevedobarroso25}
{Acevedo Barroso}, J.~A., {O'Riordan}, C.~M., {Cl{\'e}ment}, B., {et~al.} 2025,
  \aap, 697, A14

\bibitem[{{Andika} {et~al.}(2025){Andika}, {Schuldt}, {Suyu}, {Bag},
  {Ca{\~n}ameras}, {Melo}, {Grillo}, \& {Chan}}]{andika25}
{Andika}, I.~T., {Schuldt}, S., {Suyu}, S.~H., {et~al.} 2025, \aap, 694, A227

\bibitem[{{Andika} {et~al.}(2023){Andika}, {Suyu}, {Ca{\~n}ameras}, {Melo},
  {Schuldt}, {Shu}, {Eilers}, {Jaelani}, \& {Yue}}]{andika23}
{Andika}, I.~T., {Suyu}, S.~H., {Ca{\~n}ameras}, R., {et~al.} 2023, \aap, 678,
  A103

\bibitem[{{Astropy Collaboration} {et~al.}(2013){Astropy Collaboration},
  {Robitaille}, {Tollerud}, {Greenfield}, {Droettboom}, {Bray}, {Aldcroft},
  {Davis}, {Ginsburg}, {Price-Whelan}, {Kerzendorf}, {Conley}, {Crighton},
  {Barbary}, {Muna}, {Ferguson}, {Grollier}, {Parikh}, {Nair}, {Unther},
  {Deil}, {Woillez}, {Conseil}, {Kramer}, {Turner}, {Singer}, {Fox}, {Weaver},
  {Zabalza}, {Edwards}, {Azalee Bostroem}, {Burke}, {Casey}, {Crawford},
  {Dencheva}, {Ely}, {Jenness}, {Labrie}, {Lim}, {Pierfederici}, {Pontzen},
  {Ptak}, {Refsdal}, {Servillat}, \& {Streicher}}]{astropy1}
{Astropy Collaboration}, {Robitaille}, T.~P., {Tollerud}, E.~J., {et~al.} 2013,
  \aap, 558, A33

\bibitem[{{Bertin} \& {Arnouts}(1996)}]{bertin96}
{Bertin}, E. \& {Arnouts}, S. 1996, \aaps, 117, 393

\bibitem[{{Bolton} {et~al.}(2004){Bolton}, {Burles}, {Schlegel}, {Eisenstein},
  \& {Brinkmann}}]{bolton04}
{Bolton}, A.~S., {Burles}, S., {Schlegel}, D.~J., {Eisenstein}, D.~J., \&
  {Brinkmann}, J. 2004, \aj, 127, 1860

\bibitem[{{Bolton} {et~al.}(2008){Bolton}, {Treu}, {Koopmans}, {Gavazzi},
  {Moustakas}, {Burles}, {Schlegel}, \& {Wayth}}]{bolton08}
{Bolton}, A.~S., {Treu}, T., {Koopmans}, L. V.~E., {et~al.} 2008, \apj, 684,
  248

\bibitem[{{Bonvin} {et~al.}(2017){Bonvin}, {Courbin}, {Suyu}, {Marshall},
  {Rusu}, {Sluse}, {Tewes}, {Wong}, {Collett}, {Fassnacht}, {Treu}, {Auger},
  {Hilbert}, {Koopmans}, {Meylan}, {Rumbaugh}, {Sonnenfeld}, \&
  {Spiniello}}]{bonvin17}
{Bonvin}, V., {Courbin}, F., {Suyu}, S.~H., {et~al.} 2017, \mnras, 465, 4914

\bibitem[{{Ca{\~n}ameras} {et~al.}(2024){Ca{\~n}ameras}, {Schuldt}, {Shu},
  {Suyu}, {Taubenberger}, {Andika}, {Bag}, {Inoue}, {Jaelani},
  {Leal-Taix{\'e}}, {Meinhardt}, {Melo}, \& {More}}]{canameras24}
{Ca{\~n}ameras}, R., {Schuldt}, S., {Shu}, Y., {et~al.} 2024, \aap, 692, A72

\bibitem[{{Ca{\~n}ameras} {et~al.}(2021){Ca{\~n}ameras}, {Schuldt}, {Shu},
  {Suyu}, {Taubenberger}, {Meinhardt}, {Leal-Taix{\'e}}, {Chao}, {Inoue},
  {Jaelani}, \& {More}}]{canameras21b}
{Ca{\~n}ameras}, R., {Schuldt}, S., {Shu}, Y., {et~al.} 2021, \aap, 653, L6

\bibitem[{{Ca{\~n}ameras} {et~al.}(2020){Ca{\~n}ameras}, {Schuldt}, {Suyu},
  {Taubenberger}, {Meinhardt}, {Leal-Taix{\'e}}, {Lemon}, {Rojas}, \&
  {Savary}}]{canameras20}
{Ca{\~n}ameras}, R., {Schuldt}, S., {Suyu}, S.~H., {et~al.} 2020, \aap, 644,
  A163

\bibitem[{{Cabanac} {et~al.}(2007){Cabanac}, {Alard}, {Dantel-Fort}, {Fort},
  {Gavazzi}, {Gomez}, {Kneib}, {Le F{\`e}vre}, {Mellier}, {Pello}, {Soucail},
  {Sygnet}, \& {Valls-Gabaud}}]{cabanac07}
{Cabanac}, R.~A., {Alard}, C., {Dantel-Fort}, M., {et~al.} 2007, \aap, 461, 813

\bibitem[{{Caminha} {et~al.}(2019){Caminha}, {Rosati}, {Grillo}, {Rosani},
  {Caputi}, {Meneghetti}, {Mercurio}, {Balestra}, {Bergamini}, {Biviano},
  {Nonino}, {Umetsu}, {Vanzella}, {Annunziatella}, {Broadhurst},
  {Delgado-Correal}, {Demarco}, {Koekemoer}, {Lombardi}, {Maier}, {Verdugo}, \&
  {Zitrin}}]{caminha19}
{Caminha}, G.~B., {Rosati}, P., {Grillo}, C., {et~al.} 2019, \aap, 632, A36

\bibitem[{{Cao} {et~al.}(2020){Cao}, {Li}, {Shu}, {Mao}, {Kneib}, \&
  {Gao}}]{cao20}
{Cao}, X., {Li}, R., {Shu}, Y., {et~al.} 2020, \mnras, 499, 3610

\bibitem[{{Chan} {et~al.}(2020){Chan}, {Suyu}, {Sonnenfeld}, {Jaelani}, {More},
  {Yonehara}, {Kubota}, {Coupon}, {Lee}, {Oguri}, {Rusu}, \& {Wong}}]{chan20}
{Chan}, J. H.~H., {Suyu}, S.~H., {Sonnenfeld}, A., {et~al.} 2020, \aap, 636,
  A87

\bibitem[{{Courbin} {et~al.}(2018){Courbin}, {Bonvin}, {Buckley-Geer},
  {Fassnacht}, {Frieman}, {Lin}, {Marshall}, {Suyu}, {Treu}, {Anguita},
  {Motta}, {Meylan}, {Paic}, {Tewes}, {Agnello}, {Chao}, {Chijani}, {Gilman},
  {Rojas}, {Williams}, {Hempel}, {Kim}, {Lachaume}, {Rabus}, {Abbott}, {Allam},
  {Annis}, {Banerji}, {Bechtol}, {Benoit-L{\'e}vy}, {Brooks}, {Burke}, {Carnero
  Rosell}, {Carrasco Kind}, {Carretero}, {D'Andrea}, {da Costa}, {Davis},
  {DePoy}, {Desai}, {Flaugher}, {Fosalba}, {Garc{\'\i}a-Bellido}, {Gaztanaga},
  {Goldstein}, {Gruen}, {Gruendl}, {Gschwend}, {Gutierrez}, {Honscheid},
  {James}, {Kuehn}, {Kuhlmann}, {Kuropatkin}, {Lahav}, {Lima}, {Maia}, {March},
  {Marshall}, {McMahon}, {Menanteau}, {Miquel}, {Nord}, {Plazas}, {Sanchez},
  {Scarpine}, {Schindler}, {Schubnell}, {Sevilla-Noarbe}, {Smith},
  {Soares-Santos}, {Sobreira}, {Suchyta}, {Tarle}, {Tucker}, {Walker}, \&
  {Wester}}]{courbin18}
{Courbin}, F., {Bonvin}, V., {Buckley-Geer}, E., {et~al.} 2018, \aap, 609, A71

\bibitem[{{Diehl} {et~al.}(2017){Diehl}, {Buckley-Geer}, {Lindgren}, {Nord},
  {Gaitsch}, {Gaitsch}, {Lin}, {Allam}, {Collett}, {Furlanetto}, {Gill},
  {More}, {Nightingale}, {Odden}, {Pellico}, {Tucker}, {da Costa}, {Fausti
  Neto}, {Kuropatkin}, {Soares-Santos}, {Welch}, {Zhang}, {Frieman}, {Abdalla},
  {Annis}, {Benoit-L{\'e}vy}, {Bertin}, {Brooks}, {Burke}, {Carnero Rosell},
  {Carrasco Kind}, {Carretero}, {Cunha}, {D'Andrea}, {Desai}, {Dietrich},
  {Drlica-Wagner}, {Evrard}, {Finley}, {Flaugher}, {Garc{\'\i}a-Bellido},
  {Gerdes}, {Goldstein}, {Gruen}, {Gruendl}, {Gschwend}, {Gutierrez}, {James},
  {Kuehn}, {Kuhlmann}, {Lahav}, {Li}, {Lima}, {Maia}, {Marshall}, {Menanteau},
  {Miquel}, {Nichol}, {Nugent}, {Ogando}, {Plazas}, {Reil}, {Romer}, {Sako},
  {Sanchez}, {Santiago}, {Scarpine}, {Schindler}, {Schubnell},
  {Sevilla-Noarbe}, {Sheldon}, {Smith}, {Sobreira}, {Suchyta}, {Swanson},
  {Tarle}, {Thomas}, {Walker}, \& {DES Collaboration}}]{diehl17}
{Diehl}, H.~T., {Buckley-Geer}, E.~J., {Lindgren}, K.~A., {et~al.} 2017, \apjs,
  232, 15

\bibitem[{{Enzi} {et~al.}(2025){Enzi}, {Krawczyk}, {Ballard}, \&
  {Collett}}]{enzi25}
{Enzi}, W. J.~R., {Krawczyk}, C.~M., {Ballard}, D.~J., \& {Collett}, T.~E.
  2025, \mnras, 540, 247

\bibitem[{{Ertl} {et~al.}(2024){Ertl}, {Schuldt}, {Suyu}, {Schechter},
  {Halkola}, \& {Wagner}}]{ertl24_HE0230}
{Ertl}, S., {Schuldt}, S., {Suyu}, S.~H., {et~al.} 2024, \aap, 685, A15

\bibitem[{{Ertl} {et~al.}(2025){Ertl}, {Suyu}, {Schuldt}, {Granata}, {Grillo},
  {Caminha}, {Acebron}, {Bergamini}, {Ca{\~n}ameras}, {Cha}, {Diego}, {Foo},
  {Frye}, {Fudamoto}, {Halkola}, {Jee}, {Kamieneski}, {Koekemoer}, {Meena},
  {Nishida}, {Oguri}, {Pierel}, {Rosati}, {Tortorelli}, {Wang}, \&
  {Zitrin}}]{ertl25_encore}
{Ertl}, S., {Suyu}, S.~H., {Schuldt}, S., {et~al.} 2025, arXiv e-prints,
  arXiv:2503.09718

\bibitem[{{Euclid Collaboration} {et~al.}(2025{\natexlab{a}}){Euclid
  Collaboration}, {Holloway}, {Verma}, {Walmsley}, {Marshall}, {More},
  {Collett}, {Lines}, {Leuzzi}, {Manj{\'o}n-Garc{\'\i}a}, {Vincken}, {Wilde},
  {Pearce-Casey}, {Andika}, {Acevedo Barroso}, {Li}, {Melo}, {Metcalf},
  {Rojas}, {Cl{\'e}ment}, {Degaudenzi}, {Courbin}, {Despali}, {Gavazzi},
  {Schuldt}, {Nagam}, {Sluse}, {Tortora}, {Dom{\'\i}nguez S{\'a}nchez},
  {Finner}, {Galan}, {Giocoli}, {Guzzo}, {Hogg}, {Jahnke}, {Kruk}, {Mahler},
  {Millon}, {Nugent}, {Pearson}, {Ecker}, {Sainz de Murieta}, {Scarlata},
  {Serjeant}, {Sonnenfeld}, {Spiniello}, {Thai}, {Ulivi}, {Weisenbach},
  {Zumalacarregui}, {Aghanim}, {Altieri}, {Amara}, {Andreon}, {Auricchio},
  {Aussel}, {Baccigalupi}, {Baldi}, {Balestra}, {Bardelli}, {Battaglia},
  {Bender}, {Biviano}, {Bonchi}, {Branchini}, {Brescia}, {Brinchmann},
  {Camera}, {Ca{\~n}as-Herrera}, {Capobianco}, {Carbone}, {Cardone},
  {Carretero}, {Castellano}, {Castignani}, {Cavuoti}, {Chambers}, {Cimatti},
  {Colodro-Conde}, {Congedo}, {Conselice}, {Conversi}, {Copin}, {Courtois},
  {Cropper}, {Da Silva}, {De Lucia}, {Di Giorgio}, {Dolding}, {Dole}, {Dubath},
  {Duncan}, {Dupac}, {Dusini}, {Ealet}, {Escoffier}, {Farina}, {Farinelli},
  {Faustini}, {Ferriol}, {Finelli}, {Fotopoulou}, {Frailis}, {Franceschi},
  {Fumana}, {Galeotta}, {George}, {Gillis}, {G{\'o}mez-Alvarez},
  {Gracia-Carpio}, {Granett}, {Grazian}, {Grupp}, {Haugan}, {Hoar}, {Holmes},
  {Hormuth}, {Hornstrup}, {Hudelot}, {Jhabvala}, {Joachimi}, {Keih{\"a}nen},
  {Kermiche}, {Kiessling}, {Kubik}, {K{\"u}mmel}, {Kunz}, {Kurki-Suonio}, {Le
  Boulc'h}, {Le Brun}, {Le Mignant}, {Ligori}, {Lilje}, {Lindholm}, {Lloro},
  {Mainetti}, {Maino}, {Maiorano}, {Mansutti}, {Marcin}, {Marggraf},
  {Martinelli}, {Martinet}, {Marulli}, {Massey}, {Maurogordato}, {Medinaceli},
  {Mei}, {Melchior}, {Mellier}, {Meneghetti}, {Merlin}, {Meylan}, {Mora},
  {Moresco}, {Moscardini}, {Nakajima}, {Neissner}, {Nichol}, {Niemi},
  {Nightingale}, {Padilla}, {Paltani}, {Pasian}, {Pedersen}, {Percival},
  {Pettorino}, {Pires}, {Polenta}, {Poncet}, {Popa}, {Pozzetti}, {Raison},
  {Rebolo}, {Renzi}, {Rhodes}, {Riccio}, {Romelli}, {Roncarelli}, {Saglia},
  {Sakr}, {Sapone}, {Sartoris}, {Schewtschenko}, {Schneider}, {Secroun},
  {Seidel}, {Serrano}, {Simon}, {Sirignano}, {Sirri}, {Stanco}, {Steinwagner},
  {Tallada-Cresp{\'\i}}, {Taylor}, {Tereno}, {Toft}, {Toledo-Moreo},
  {Torradeflot}, \& {Tutusaus}}]{holloway25_Q1}
{Euclid Collaboration}, {Holloway}, P., {Verma}, A., {et~al.}
  2025{\natexlab{a}}, arXiv e-prints, arXiv:2503.15328

\bibitem[{{Euclid Collaboration} {et~al.}(2025{\natexlab{b}}){Euclid
  Collaboration}, {Walmsley}, {Holloway}, {Lines}, {Rojas}, {Collett}, {Verma},
  {Li}, {Nightingale}, {Despali}, {Schuldt}, {Gavazzi}, {Melo}, {Metcalf},
  {Andika}, {Leuzzi}, {Manj{\'o}n-Garc{\'\i}a}, {Pearce-Casey}, {Vincken},
  {Wilde}, {Busillo}, {Tortora}, {Acevedo Barroso}, {Dole}, {Ecker}, {Pearson},
  {Marshall}, {More}, {Saifollahi}, {Gracia-Carpio}, {Baeten}, {Cornen},
  {Johnson}, {Macmillan}, {Kruk}, {Remmelgas}, {Cl{\'e}ment}, {Degaudenzi},
  {Courbin}, {Bovy}, {Casas}, {Dannerbauer}, {Diego}, {Finner}, {Galan},
  {Giocoli}, {Hogg}, {Jahnke}, {Katona}, {Kov{\'a}cs}, {De Leo}, {Mahler},
  {Millon}, {Nagam}, {Nugent}, {Sainz de Murieta}, {O'Riordan}, {Sluse},
  {Sonnenfeld}, {Spiniello}, {Serjeant}, {Thai}, {Ulivi}, {Walth},
  {Weisenbach}, {Zumalacarregui}, {Aghanim}, {Altieri}, {Amara}, {Andreon},
  {Auricchio}, {Aussel}, {Baccigalupi}, {Baldi}, {Balestra}, {Bardelli},
  {Battaglia}, {Bernardeau}, {Biviano}, {Bonchi}, {Bonino}, {Branchini},
  {Brescia}, {Brinchmann}, {Camera}, {Ca{\~n}as-Herrera}, {Capobianco},
  {Carbone}, {Cardone}, {Carretero}, {Castander}, {Castellano}, {Castignani},
  {Cavuoti}, {Chambers}, {Cimatti}, {Colodro-Conde}, {Congedo}, {Conselice},
  {Conversi}, {Copin}, {Corcione}, {Courtois}, {Cropper}, {Da Silva}, {De
  Lucia}, {Di Giorgio}, {Dolding}, {Dubath}, {Duncan}, {Dupac}, {Ealet},
  {Escoffier}, {Fabricius}, {Farina}, {Farinelli}, {Faustini}, {Finelli},
  {Fotopoulou}, {Frailis}, {Franceschi}, {Fumana}, {Galeotta}, {George},
  {Gillard}, {Gillis}, {G{\'o}mez-Alvarez}, {Granett}, {Grazian}, {Grupp},
  {Guzzo}, {Gwyn}, {Haugan}, {Hoekstra}, {Holmes}, {Hook}, {Hormuth},
  {Hornstrup}, {Hudelot}, {Jhabvala}, {Joachimi}, {Keih{\"a}nen}, {Kermiche},
  {Kiessling}, {Kubik}, {K{\"u}mmel}, {Kunz}, {Kurki-Suonio}, {Lahav}, {Le
  Boulc'h}, {Le Brun}, {Le Mignant}, {Ligori}, {Lilje}, {Lindholm}, {Lloro},
  {Mainetti}, {Maino}, {Maiorano}, {Mansutti}, {Marcin}, {Marggraf},
  {Martinelli}, {Martinet}, {Marulli}, {Massey}, {Maurogordato}, {McCracken},
  {Medinaceli}, {Mei}, {Mellier}, {Meneghetti}, {Merlin}, {Meylan}, {Mora},
  {Moresco}, {Moscardini}, {Nakajima}, {Neissner}, {Nichol}, {Niemi},
  {Padilla}, {Paltani}, {Pasian}, {Pedersen}, {Percival}, {Pettorino}, {Pires},
  {Polenta}, {Poncet}, {Popa}, {Pozzetti}, {Raison}, {Rebolo}, {Renzi},
  {Rhodes}, {Riccio}, {Romelli}, {Roncarelli}, \& {Saglia}}]{walmsley25}
{Euclid Collaboration}, {Walmsley}, M., {Holloway}, P., {et~al.}
  2025{\natexlab{b}}, arXiv e-prints, arXiv:2503.15324

\bibitem[{{Frye} {et~al.}(2024){Frye}, {Pascale}, {Pierel}, {Chen}, {Foo},
  {Leimbach}, {Garuda}, {Cohen}, {Kamieneski}, {Windhorst}, {Koekemoer},
  {Kelly}, {Summers}, {Engesser}, {Liu}, {Furtak}, {Polletta}, {Harrington},
  {Willner}, {Diego}, {Jansen}, {Coe}, {Conselice}, {Dai}, {Dole}, {D'Silva},
  {Driver}, {Grogin}, {Marshall}, {Meena}, {Nonino}, {Ortiz}, {Pirzkal},
  {Robotham}, {Ryan}, {Strolger}, {Tompkins}, {Willmer}, {Yan}, {Yun}, \&
  {Zitrin}}]{frye24}
{Frye}, B.~L., {Pascale}, M., {Pierel}, J., {et~al.} 2024, \apj, 961, 171

\bibitem[{{Gavazzi} {et~al.}(2014){Gavazzi}, {Marshall}, {Treu}, \&
  {Sonnenfeld}}]{gavazzi14}
{Gavazzi}, R., {Marshall}, P.~J., {Treu}, T., \& {Sonnenfeld}, A. 2014, \apj,
  785, 144

\bibitem[{{Gong} {et~al.}(2019){Gong}, {Liu}, {Cao}, {Chen}, {Fan}, {Li}, {Li},
  {Li}, {Zhang}, \& {Zhan}}]{gong19}
{Gong}, Y., {Liu}, X., {Cao}, Y., {et~al.} 2019, \apj, 883, 203

\bibitem[{{Granata} {et~al.}(2025){Granata}, {Caminha}, {Ertl}, {Grillo},
  {Schuldt}, {Suyu}, {Acebron}, {Bergamini}, {Ca{\~n}ameras}, {Koekemoer},
  {Rosati}, \& {Taubenberger}}]{granata25_encore}
{Granata}, G., {Caminha}, G.~B., {Ertl}, S., {et~al.} 2025, \aap, 697, A94

\bibitem[{{Grespan} {et~al.}(2024){Grespan}, {Thuruthipilly}, {Pollo},
  {Lochner}, {Biesiada}, \& {Etsebeth}}]{grespan24}
{Grespan}, M., {Thuruthipilly}, H., {Pollo}, A., {et~al.} 2024, \aap, 688, A34

\bibitem[{{Grillo} {et~al.}(2024){Grillo}, {Pagano}, {Rosati}, \&
  {Suyu}}]{grillo24}
{Grillo}, C., {Pagano}, L., {Rosati}, P., \& {Suyu}, S.~H. 2024, \aap, 684, L23

\bibitem[{{Grillo} {et~al.}(2018){Grillo}, {Rosati}, {Suyu}, {Balestra},
  {Caminha}, {Halkola}, {Kelly}, {Lombardi}, {Mercurio}, {Rodney}, \&
  {Treu}}]{grillo18}
{Grillo}, C., {Rosati}, P., {Suyu}, S.~H., {et~al.} 2018, \apj, 860, 94

\bibitem[{Harris {et~al.}(2020)Harris, Millman, van~der Walt, Gommers,
  Virtanen, Cournapeau, Wieser, Taylor, Berg, Smith, Kern, Picus, Hoyer, van
  Kerkwijk, Brett, Haldane, Fernández~del Río, Wiebe, Peterson,
  Gérard-Marchant, Sheppard, Reddy, Weckesser, Abbasi, Gohlke, \&
  Oliphant}]{numpy2}
Harris, C.~R., Millman, K.~J., van~der Walt, S.~J., {et~al.} 2020, Nature, 585,
  357–362

\bibitem[{He {et~al.}(2015)He, Zhang, Ren, \& Sun}]{he16}
He, K., Zhang, X., Ren, S., \& Sun, J. 2015, Deep Residual Learning for Image
  Recognition

\bibitem[{{Holloway} {et~al.}(2024){Holloway}, {Marshall}, {Verma}, {More},
  {Ca{\~n}ameras}, {Jaelani}, {Ishida}, \& {Wong}}]{holloway24}
{Holloway}, P., {Marshall}, P.~J., {Verma}, A., {et~al.} 2024, \mnras, 530,
  1297

\bibitem[{{Holwerda} {et~al.}(2015){Holwerda}, {Baldry}, {Alpaslan}, {Bauer},
  {Bland-Hawthorn}, {Brough}, {Brown}, {Cluver}, {Conselice}, {Driver},
  {Hopkins}, {Jones}, {L{\'o}pez-S{\'a}nchez}, {Loveday}, {Meyer}, \&
  {Moffett}}]{holwerda15}
{Holwerda}, B.~W., {Baldry}, I.~K., {Alpaslan}, M., {et~al.} 2015, \mnras, 449,
  4277

\bibitem[{{Hsieh} \& {Yee}(2014)}]{hsieh14}
{Hsieh}, B.~C. \& {Yee}, H.~K.~C. 2014, \apj, 792, 102

\bibitem[{{Huang} {et~al.}(2021){Huang}, {Storfer}, {Gu}, {Ravi}, {Pilon},
  {Sheu}, {Venguswamy}, {Banka}, {Dey}, {Landriau}, {Lang}, {Meisner},
  {Moustakas}, {Myers}, {Sajith}, {Schlafly}, \& {Schlegel}}]{huang21}
{Huang}, X., {Storfer}, C., {Gu}, A., {et~al.} 2021, \apj, 909, 27

\bibitem[{{Huang} {et~al.}(2020){Huang}, {Storfer}, {Ravi}, {Pilon}, {Domingo},
  {Schlegel}, {Bailey}, {Dey}, {Gupta}, {Herrera}, {Juneau}, {Landriau},
  {Lang}, {Meisner}, {Moustakas}, {Myers}, {Schlafly}, {Valdes}, {Weaver},
  {Yang}, \& {Y{\`e}che}}]{huang20}
{Huang}, X., {Storfer}, C., {Ravi}, V., {et~al.} 2020, \apj, 894, 78

\bibitem[{Hunter(2007)}]{matplotlib}
Hunter, J.~D. 2007, Computing in Science \& Engineering, 9, 90

\bibitem[{{Inami} {et~al.}(2017){Inami}, {Bacon}, {Brinchmann}, {Richard},
  {Contini}, {Conseil}, {Hamer}, {Akhlaghi}, {Bouch{\'e}}, {Cl{\'e}ment},
  {Desprez}, {Drake}, {Hashimoto}, {Leclercq}, {Maseda}, {Michel-Dansac},
  {Paalvast}, {Tresse}, {Ventou}, {Kollatschny}, {Boogaard}, {Finley},
  {Marino}, {Schaye}, \& {Wisotzki}}]{inami17}
{Inami}, H., {Bacon}, R., {Brinchmann}, J., {et~al.} 2017, \aap, 608, A2

\bibitem[{{Ivezic} {et~al.}(2008){Ivezic}, {Axelrod}, {Brandt}, {Burke},
  {Claver}, {Connolly}, {Cook}, {Gee}, {Gilmore}, {Jacoby}, {Jones}, {Kahn},
  {Kantor}, {Krabbendam}, {Lupton}, {Monet}, {Pinto}, {Saha}, {Schalk},
  {Schneider}, {Strauss}, {Stubbs}, {Sweeney}, {Szalay}, {Thaler}, {Tyson}, \&
  {LSST Collaboration}}]{ivezic08}
{Ivezic}, Z., {Axelrod}, T., {Brandt}, W.~N., {et~al.} 2008, Serbian
  Astronomical Journal, 176, 1

\bibitem[{{Ivezi{\'c}} {et~al.}(2019){Ivezi{\'c}}, {Kahn}, {Tyson}, {Abel},
  {Acosta}, {Allsman}, {Alonso}, {AlSayyad}, {Anderson}, {Andrew}, {Angel},
  {Angeli}, {Ansari}, {Antilogus}, {Araujo}, {Armstrong}, {Arndt}, {Astier},
  {Aubourg}, {Auza}, {Axelrod}, {Bard}, {Barr}, {Barrau}, {Bartlett}, {Bauer},
  {Bauman}, {Baumont}, {Bechtol}, {Bechtol}, {Becker}, {Becla}, {Beldica},
  {Bellavia}, {Bianco}, {Biswas}, {Blanc}, {Blazek}, {Blandford}, {Bloom},
  {Bogart}, {Bond}, {Booth}, {Borgland}, {Borne}, {Bosch}, {Boutigny},
  {Brackett}, {Bradshaw}, {Brandt}, {Brown}, {Bullock}, {Burchat}, {Burke},
  {Cagnoli}, {Calabrese}, {Callahan}, {Callen}, {Carlin}, {Carlson},
  {Chandrasekharan}, {Charles-Emerson}, {Chesley}, {Cheu}, {Chiang}, {Chiang},
  {Chirino}, {Chow}, {Ciardi}, {Claver}, {Cohen-Tanugi}, {Cockrum}, {Coles},
  {Connolly}, {Cook}, {Cooray}, {Covey}, {Cribbs}, {Cui}, {Cutri}, {Daly},
  {Daniel}, {Daruich}, {Daubard}, {Daues}, {Dawson}, {Delgado}, {Dellapenna},
  {de Peyster}, {de Val-Borro}, {Digel}, {Doherty}, {Dubois},
  {Dubois-Felsmann}, {Durech}, {Economou}, {Eifler}, {Eracleous}, {Emmons},
  {Fausti Neto}, {Ferguson}, {Figueroa}, {Fisher-Levine}, {Focke}, {Foss},
  {Frank}, {Freemon}, {Gangler}, {Gawiser}, {Geary}, {Gee}, {Geha}, {Gessner},
  {Gibson}, {Gilmore}, {Glanzman}, {Glick}, {Goldina}, {Goldstein}, {Goodenow},
  {Graham}, {Gressler}, {Gris}, {Guy}, {Guyonnet}, {Haller}, {Harris},
  {Hascall}, {Haupt}, {Hernandez}, {Herrmann}, {Hileman}, {Hoblitt}, {Hodgson},
  {Hogan}, {Howard}, {Huang}, {Huffer}, {Ingraham}, {Innes}, {Jacoby}, {Jain},
  {Jammes}, {Jee}, {Jenness}, {Jernigan}, {Jevremovi{\'c}}, {Johns}, {Johnson},
  {Johnson}, {Jones}, {Juramy-Gilles}, {Juri{\'c}}, {Kalirai}, {Kallivayalil},
  {Kalmbach}, {Kantor}, {Karst}, {Kasliwal}, {Kelly}, {Kessler}, {Kinnison},
  {Kirkby}, {Knox}, {Kotov}, {Krabbendam}, {Krughoff}, {Kub{\'a}nek},
  {Kuczewski}, {Kulkarni}, {Ku}, {Kurita}, {Lage}, {Lambert}, {Lange},
  {Langton}, {Le Guillou}, {Levine}, {Liang}, {Lim}, {Lintott}, {Long},
  {Lopez}, {Lotz}, {Lupton}, {Lust}, {MacArthur}, {Mahabal}, {Mandelbaum},
  {Markiewicz}, {Marsh}, {Marshall}, {Marshall}, {May}, {McKercher}, {McQueen},
  {Meyers}, {Migliore}, {Miller}, \& {Mills}}]{ivezic19}
{Ivezi{\'c}}, {\v{Z}}., {Kahn}, S.~M., {Tyson}, J.~A., {et~al.} 2019, \apj,
  873, 111

\bibitem[{{Jacobs} {et~al.}(2019){Jacobs}, {Collett}, {Glazebrook},
  {Buckley-Geer}, {Diehl}, {Lin}, {McCarthy}, {Qin}, {Odden}, {Caso Escudero},
  {Dial}, {Yung}, {Gaitsch}, {Pellico}, {Lindgren}, {Abbott}, {Annis}, {Avila},
  {Brooks}, {Burke}, {Carnero Rosell}, {Carrasco Kind}, {Carretero}, {da
  Costa}, {De Vicente}, {Fosalba}, {Frieman}, {Garc{\'\i}a-Bellido},
  {Gaztanaga}, {Goldstein}, {Gruen}, {Gruendl}, {Gschwend}, {Hollowood},
  {Honscheid}, {Hoyle}, {James}, {Krause}, {Kuropatkin}, {Lahav}, {Lima},
  {Maia}, {Marshall}, {Miquel}, {Plazas}, {Roodman}, {Sanchez}, {Scarpine},
  {Serrano}, {Sevilla-Noarbe}, {Smith}, {Sobreira}, {Suchyta}, {Swanson},
  {Tarle}, {Vikram}, {Walker}, {Zhang}, \& {DES Collaboration}}]{jacobs19}
{Jacobs}, C., {Collett}, T., {Glazebrook}, K., {et~al.} 2019, \apjs, 243, 17

\bibitem[{{Jacobs} {et~al.}(2017){Jacobs}, {Glazebrook}, {Collett}, {More}, \&
  {McCarthy}}]{jacobs17}
{Jacobs}, C., {Glazebrook}, K., {Collett}, T., {More}, A., \& {McCarthy}, C.
  2017, \mnras, 471, 167

\bibitem[{{Jaelani} {et~al.}(2020){Jaelani}, {More}, {Sonnenfeld}, {Oguri},
  {Rusu}, {Wong}, {Chan}, {Suyu}, {Kayo}, {Lee}, \& {Inoue}}]{jaelani20b}
{Jaelani}, A.~T., {More}, A., {Sonnenfeld}, A., {et~al.} 2020, \mnras, 494,
  3156

\bibitem[{{Jaelani} {et~al.}(2024){Jaelani}, {More}, {Wong}, {Inoue}, {Chao},
  {Premadi}, \& {Ca{\~n}ameras}}]{jaelani24}
{Jaelani}, A.~T., {More}, A., {Wong}, K.~C., {et~al.} 2024, \mnras, 535, 1625

\bibitem[{{Kelly} {et~al.}(2023){Kelly}, {Rodney}, {Treu}, {Oguri}, {Chen},
  {Zitrin}, {Birrer}, {Bonvin}, {Dessart}, {Diego}, {Filippenko}, {Foley},
  {Gilman}, {Hjorth}, {Jauzac}, {Mandel}, {Millon}, {Pierel}, {Sharon},
  {Thorp}, {Williams}, {Broadhurst}, {Dressler}, {Graur}, {Jha}, {McCully},
  {Postman}, {Schmidt}, {Tucker}, \& {von der Linden}}]{kelly23}
{Kelly}, P.~L., {Rodney}, S., {Treu}, T., {et~al.} 2023, Science, 380, abh1322

\bibitem[{{Lange} {et~al.}(2025){Lange}, {Amvrosiadis}, {Nightingale}, {He},
  {Frenk}, {Robertson}, {Cole}, {Massey}, {Cao}, {Li}, \& {Wang}}]{lange25}
{Lange}, S.~C., {Amvrosiadis}, A., {Nightingale}, J.~W., {et~al.} 2025, \mnras,
  539, 704

\bibitem[{{Lanusse} {et~al.}(2018){Lanusse}, {Ma}, {Li}, {Collett}, {Li},
  {Ravanbakhsh}, {Mandelbaum}, \& {P{\'o}czos}}]{lanusse18}
{Lanusse}, F., {Ma}, Q., {Li}, N., {et~al.} 2018, \mnras, 473, 3895

\bibitem[{{Laureijs} {et~al.}(2011){Laureijs}, {Amiaux}, {Arduini},
  {Augu{\`e}res}, {Brinchmann}, {Cole}, {Cropper}, {Dabin}, {Duvet}, {Ealet},
  {Garilli}, {Gondoin}, {Guzzo}, {Hoar}, {Hoekstra}, {Holmes}, {Kitching},
  {Maciaszek}, {Mellier}, {Pasian}, {Percival}, {Rhodes}, {Saavedra Criado},
  {Sauvage}, {Scaramella}, {Valenziano}, {Warren}, {Bender}, {Castander},
  {Cimatti}, {Le F{\`e}vre}, {Kurki-Suonio}, {Levi}, {Lilje}, {Meylan},
  {Nichol}, {Pedersen}, {Popa}, {Rebolo Lopez}, {Rix}, {Rottgering},
  {Zeilinger}, {Grupp}, {Hudelot}, {Massey}, {Meneghetti}, {Miller}, {Paltani},
  {Paulin-Henriksson}, {Pires}, {Saxton}, {Schrabback}, {Seidel}, {Walsh},
  {Aghanim}, {Amendola}, {Bartlett}, {Baccigalupi}, {Beaulieu}, {Benabed},
  {Cuby}, {Elbaz}, {Fosalba}, {Gavazzi}, {Helmi}, {Hook}, {Irwin}, {Kneib},
  {Kunz}, {Mannucci}, {Moscardini}, {Tao}, {Teyssier}, {Weller}, {Zamorani},
  {Zapatero Osorio}, {Boulade}, {Foumond}, {Di Giorgio}, {Guttridge}, {James},
  {Kemp}, {Martignac}, {Spencer}, {Walton}, {Bl{\"u}mchen}, {Bonoli},
  {Bortoletto}, {Cerna}, {Corcione}, {Fabron}, {Jahnke}, {Ligori}, {Madrid},
  {Martin}, {Morgante}, {Pamplona}, {Prieto}, {Riva}, {Toledo}, {Trifoglio},
  {Zerbi}, {Abdalla}, {Douspis}, {Grenet}, {Borgani}, {Bouwens}, {Courbin},
  {Delouis}, {Dubath}, {Fontana}, {Frailis}, {Grazian}, {Koppenh{\"o}fer},
  {Mansutti}, {Melchior}, {Mignoli}, {Mohr}, {Neissner}, {Noddle}, {Poncet},
  {Scodeggio}, {Serrano}, {Shane}, {Starck}, {Surace}, {Taylor},
  {Verdoes-Kleijn}, {Vuerli}, {Williams}, {Zacchei}, {Altieri}, {Escudero
  Sanz}, {Kohley}, {Oosterbroek}, {Astier}, {Bacon}, {Bardelli}, {Baugh},
  {Bellagamba}, {Benoist}, {Bianchi}, {Biviano}, {Branchini}, {Carbone},
  {Cardone}, {Clements}, {Colombi}, {Conselice}, {Cresci}, {Deacon}, {Dunlop},
  {Fedeli}, {Fontanot}, {Franzetti}, {Giocoli}, {Garcia-Bellido}, {Gow},
  {Heavens}, {Hewett}, {Heymans}, {Holland}, {Huang}, {Ilbert}, {Joachimi},
  {Jennins}, {Kerins}, {Kiessling}, {Kirk}, {Kotak}, {Krause}, {Lahav}, {van
  Leeuwen}, {Lesgourgues}, {Lombardi}, {Magliocchetti}, {Maguire}, {Majerotto},
  {Maoli}, {Marulli}, {Maurogordato}, {McCracken}, {McLure}, {Melchiorri},
  {Merson}, {Moresco}, {Nonino}, {Norberg}, {Peacock}, {Pello}, {Penny},
  {Pettorino}, {Di Porto}, {Pozzetti}, {Quercellini}, {Radovich}, {Rassat},
  {Roche}, {Ronayette}, {Rossetti}, {Sartoris}, {Schneider}, {Semboloni},
  {Serjeant}, {Simpson}, {Skordis}, {Smadja}, {Smartt}, {Spano}, {Spiro},
  {Sullivan}, {Tilquin}, {Trotta}, {Verde}, {Wang}, {Williger}, {Zhao},
  {Zoubian}, \& {Zucca}}]{laureijs11}
{Laureijs}, R., {Amiaux}, J., {Arduini}, S., {et~al.} 2011, arXiv e-prints,
  arXiv:1110.3193

\bibitem[{{Li} {et~al.}(2021){Li}, {Napolitano}, {Spiniello}, {Tortora},
  {Kuijken}, {Koopmans}, {Schneider}, {Getman}, {Xie}, {Long}, {Shu},
  {Vernardos}, {Huang}, {Covone}, {Dvornik}, {Heymans}, {Hildebrandt},
  {Radovich}, \& {Wright}}]{li21c_search_KiDS}
{Li}, R., {Napolitano}, N.~R., {Spiniello}, C., {et~al.} 2021, \apj, 923, 16

\bibitem[{{Li} {et~al.}(2020){Li}, {Napolitano}, {Tortora}, {Spiniello},
  {Koopmans}, {Huang}, {Roy}, {Vernardos}, {Chatterjee}, {Giblin}, {Getman},
  {Radovich}, {Covone}, \& {Kuijken}}]{li20}
{Li}, R., {Napolitano}, N.~R., {Tortora}, C., {et~al.} 2020, \apj, 899, 30

\bibitem[{{Li} {et~al.}(2019){Li}, {Shu}, {Su}, {Feng}, {Zhang}, {Wang}, \&
  {Liu}}]{li19}
{Li}, R., {Shu}, Y., {Su}, J., {et~al.} 2019, \mnras, 482, 313

\bibitem[{{Me{\v{s}}tri{\'c}} {et~al.}(2022){Me{\v{s}}tri{\'c}}, {Vanzella},
  {Zanella}, {Castellano}, {Calura}, {Rosati}, {Bergamini}, {Mercurio},
  {Meneghetti}, {Grillo}, {Caminha}, {Nonino}, {Merlin}, {Cupani}, \&
  {Sani}}]{mestric22}
{Me{\v{s}}tri{\'c}}, U., {Vanzella}, E., {Zanella}, A., {et~al.} 2022, \mnras,
  516, 3532

\bibitem[{{Millon} {et~al.}(2020){Millon}, {Courbin}, {Bonvin}, {Paic},
  {Meylan}, {Tewes}, {Sluse}, {Magain}, {Chan}, {Galan}, {Joseph}, {Lemon},
  {Tihhonova}, {Anderson}, {Marmier}, {Chazelas}, {Lendl}, {Triaud}, \&
  {Wyttenbach}}]{millon20b}
{Millon}, M., {Courbin}, F., {Bonvin}, V., {et~al.} 2020, \aap, 640, A105

\bibitem[{{More} {et~al.}(2016){More}, {Verma}, {Marshall}, {More}, {Baeten},
  {Wilcox}, {Macmillan}, {Cornen}, {Kapadia}, {Parrish}, {Snyder}, {Davis},
  {Gavazzi}, {Lintott}, {Simpson}, {Miller}, {Smith}, {Paget}, {Saha},
  {K{\"u}ng}, \& {Collett}}]{more16b}
{More}, A., {Verma}, A., {Marshall}, P.~J., {et~al.} 2016, \mnras, 455, 1191

\bibitem[{{Morishita} {et~al.}(2024){Morishita}, {Stiavelli}, {Grillo},
  {Rosati}, {Schuldt}, {Trenti}, {Bergamini}, {Boyett}, {Chary},
  {Leethochawalit}, {Roberts-Borsani}, {Treu}, \& {Vanzella}}]{morishita24}
{Morishita}, T., {Stiavelli}, M., {Grillo}, C., {et~al.} 2024, \apj, 971, 43

\bibitem[{{Nagam} {et~al.}(2025){Nagam}, {Acevedo Barroso}, {Wilde}, {Andika},
  {Manj{\'o}n-Garc{\'\i}a}, {Pearce-Casey}, {Stern}, {Nightingale},
  {Moustakas}, {McCarthy}, {Moravec}, {Leuzzi}, {Rojas}, {Serjeant}, {Collett},
  {Matavulj}, {Walmsley}, {Cl{\'e}ment}, {Tortora}, {Gavazzi}, {Metcalf},
  {O'Riordan}, {Verdoes Kleijn}, {Koopmans}, {Valentijn}, {Busillo}, {Schuldt},
  {Courbin}, {Vernardos}, {Meneghetti}, {D{\'\i}az-S{\'a}nchez}, {Diego},
  {Ecker}, {Thai}, {Cooray}, {Courtois}, {Delchambre}, {Despali}, {Sluse},
  {Ulivi}, {Melo}, {Corcho-Caballero}, {Altieri}, {Amara}, {Andreon},
  {Auricchio}, {Aussel}, {Baccigalupi}, {Baldi}, {Balestra}, {Bardelli},
  {Battaglia}, {Bonino}, {Branchini}, {Brescia}, {Brinchmann}, {Caillat},
  {Camera}, {Capobianco}, {Carbone}, {Carretero}, {Casas}, {Castellano},
  {Castignani}, {Cavuoti}, {Cimatti}, {Colodro-Conde}, {Congedo}, {Conselice},
  {Conversi}, {Copin}, {Cropper}, {Da Silva}, {Degaudenzi}, {De Lucia}, {Di
  Giorgio}, {Dinis}, {Dubath}, {Duncan}, {Dupac}, {Dusini}, {Fabricius},
  {Farina}, {Farrens}, {Ferriol}, {Frailis}, {Franceschi}, {Fumana}, {George},
  {Gillard}, {Gillis}, {Giocoli}, {G{\'o}mez-Alvarez}, {Grazian}, {Grupp},
  {Guzzo}, {Haugan}, {Hoar}, {Holmes}, {Hook}, {Hormuth}, {Hornstrup},
  {Hudelot}, {Jahnke}, {Jhabvala}, {Joachimi}, {Keih{\"a}nen}, {Kermiche},
  {Kubik}, {Kuijken}, {K{\"u}mmel}, {Kunz}, {Kurki-Suonio}, {Laureijs}, {Le
  Mignant}, {Ligori}, {Lilje}, {Lindholm}, {Lloro}, {Mainetti}, {Maiorano},
  {Mansutti}, {Marggraf}, {Markovic}, {Martinelli}, {Martinet}, {Marulli},
  {Massey}, {Medinaceli}, {Melchior}, {Mellier}, {Merlin}, {Meylan}, {Moresco},
  {Moscardini}, {Nakajima}, {Neissner}, {Nichol}, {Niemi}, {Padilla},
  {Paltani}, {Pasian}, {Pedersen}, {Percival}, {Pettorino}, {Pires}, {Polenta},
  {Poncet}, {Popa}, {Pozzetti}, {Raison}, {Rebolo}, {Renzi}, {Rhodes},
  {Riccio}, {Romelli}, {Roncarelli}, {Rossetti}, {Saglia}, {Sakr},
  {S{\'a}nchez}, {Sapone}, {Sartoris}, {Schirmer}, {Schneider}, {Schrabback},
  {Secroun}, {Seidel}, {Serrano}, {Sirignano}, {Sirri}, {Skottfelt}, {Stanco},
  {Starck}, {Steinwagner}, {Tallada-Cresp{\'\i}}, {Tavagnacco}, {Taylor},
  {Teplitz}, {Tereno}, {Toledo-Moreo}, {Torradeflot}, {Tsyganov}, {Tutusaus},
  {Valenziano}, {Vassallo}, {Veropalumbo}, {Wang}, {Weller}, {Zacchei},
  {Zucca}, {Burigana}, {Mora}, {P{\"o}ntinen}, \& {Scottez}}]{nagam25}
{Nagam}, B.~C., {Acevedo Barroso}, J.~A., {Wilde}, J., {et~al.} 2025, arXiv
  e-prints, arXiv:2502.09802

\bibitem[{{Paraficz} {et~al.}(2016){Paraficz}, {Courbin}, {Tramacere},
  {Joseph}, {Metcalf}, {Kneib}, {Dubath}, {Droz}, {Filleul}, {Ringeisen}, \&
  {Sch{\"a}fer}}]{paraficz16}
{Paraficz}, D., {Courbin}, F., {Tramacere}, A., {et~al.} 2016, \aap, 592, A75

\bibitem[{{Pascale} {et~al.}(2025){Pascale}, {Frye}, {Pierel}, {Chen}, {Kelly},
  {Cohen}, {Windhorst}, {Riess}, {Kamieneski}, {Diego}, {Meena}, {Cha},
  {Oguri}, {Zitrin}, {Jee}, {Foo}, {Leimbach}, {Koekemoer}, {Conselice}, {Dai},
  {Goobar}, {Siebert}, {Strolger}, \& {Willner}}]{pascale25}
{Pascale}, M., {Frye}, B.~L., {Pierel}, J. D.~R., {et~al.} 2025, \apj, 979, 13

\bibitem[{{Petrillo} {et~al.}(2019){Petrillo}, {Tortora}, {Vernardos},
  {Koopmans}, {Verdoes Kleijn}, {Bilicki}, {Napolitano}, {Chatterjee},
  {Covone}, {Dvornik}, {Erben}, {Getman}, {Giblin}, {Heymans}, {de Jong},
  {Kuijken}, {Schneider}, {Shan}, {Spiniello}, \& {Wright}}]{petrillo19b}
{Petrillo}, C.~E., {Tortora}, C., {Vernardos}, G., {et~al.} 2019, \mnras, 484,
  3879

\bibitem[{{Pierel} {et~al.}(2024){Pierel}, {Newman}, {Dhawan}, {Gu}, {Joshi},
  {Li}, {Schuldt}, {Strolger}, {Suyu}, {Caminha}, {Cohen}, {Diego},
  {D{\'S}ilva}, {Ertl}, {Frye}, {Granata}, {Grillo}, {Koekemoer}, {Li},
  {Robotham}, {Summers}, {Treu}, {Windhorst}, {Zitrin}, {Agarwal}, {Agrawal},
  {Arendse}, {Belli}, {Burns}, {Ca{\~n}ameras}, {Chakrabarti}, {Chen},
  {Collett}, {Coulter}, {Ellis}, {Engesser}, {Foo}, {Fox}, {Gall}, {Garuda},
  {Gezari}, {Gomez}, {Glazebrook}, {Hjorth}, {Huang}, {Jha}, {Kamieneski},
  {Kelly}, {Larison}, {Moustakas}, {Pascale}, {P{\'e}rez-Fournon},
  {Petrushevska}, {Poidevin}, {Rest}, {Shahbandeh}, {Shajib}, {Siebert},
  {Storfer}, {Talbot}, {Wang}, {Wevers}, \& {Zenati}}]{pierel24}
{Pierel}, J.~D.~R., {Newman}, A.~B., {Dhawan}, S., {et~al.} 2024, \apjl, 967,
  L37

\bibitem[{{Planck Collaboration} {et~al.}(2020){Planck Collaboration},
  {Aghanim}, {Akrami}, {Ashdown}, {Aumont}, {Baccigalupi}, {Ballardini},
  {Banday}, {Barreiro}, {Bartolo}, {Basak}, {Battye}, {Benabed}, {Bernard},
  {Bersanelli}, {Bielewicz}, {Bock}, {Bond}, {Borrill}, {Bouchet}, {Boulanger},
  {Bucher}, {Burigana}, {Butler}, {Calabrese}, {Cardoso}, {Carron},
  {Challinor}, {Chiang}, {Chluba}, {Colombo}, {Combet}, {Contreras}, {Crill},
  {Cuttaia}, {de Bernardis}, {de Zotti}, {Delabrouille}, {Delouis}, {Di
  Valentino}, {Diego}, {Dor{\'e}}, {Douspis}, {Ducout}, {Dupac}, {Dusini},
  {Efstathiou}, {Elsner}, {En{\ss}lin}, {Eriksen}, {Fantaye}, {Farhang},
  {Fergusson}, {Fernandez-Cobos}, {Finelli}, {Forastieri}, {Frailis},
  {Fraisse}, {Franceschi}, {Frolov}, {Galeotta}, {Galli}, {Ganga},
  {G{\'e}nova-Santos}, {Gerbino}, {Ghosh}, {Gonz{\'a}lez-Nuevo}, {G{\'o}rski},
  {Gratton}, {Gruppuso}, {Gudmundsson}, {Hamann}, {Handley}, {Hansen},
  {Herranz}, {Hildebrandt}, {Hivon}, {Huang}, {Jaffe}, {Jones}, {Karakci},
  {Keih{\"a}nen}, {Keskitalo}, {Kiiveri}, {Kim}, {Kisner}, {Knox},
  {Krachmalnicoff}, {Kunz}, {Kurki-Suonio}, {Lagache}, {Lamarre}, {Lasenby},
  {Lattanzi}, {Lawrence}, {Le Jeune}, {Lemos}, {Lesgourgues}, {Levrier},
  {Lewis}, {Liguori}, {Lilje}, {Lilley}, {Lindholm}, {L{\'o}pez-Caniego},
  {Lubin}, {Ma}, {Mac{\'\i}as-P{\'e}rez}, {Maggio}, {Maino}, {Mandolesi},
  {Mangilli}, {Marcos-Caballero}, {Maris}, {Martin}, {Martinelli},
  {Mart{\'\i}nez-Gonz{\'a}lez}, {Matarrese}, {Mauri}, {McEwen}, {Meinhold},
  {Melchiorri}, {Mennella}, {Migliaccio}, {Millea}, {Mitra},
  {Miville-Desch{\^e}nes}, {Molinari}, {Montier}, {Morgante}, {Moss}, {Natoli},
  {N{\o}rgaard-Nielsen}, {Pagano}, {Paoletti}, {Partridge}, {Patanchon},
  {Peiris}, {Perrotta}, {Pettorino}, {Piacentini}, {Polastri}, {Polenta},
  {Puget}, {Rachen}, {Reinecke}, {Remazeilles}, {Renzi}, {Rocha}, {Rosset},
  {Roudier}, {Rubi{\~n}o-Mart{\'\i}n}, {Ruiz-Granados}, {Salvati}, {Sandri},
  {Savelainen}, {Scott}, {Shellard}, {Sirignano}, {Sirri}, {Spencer},
  {Sunyaev}, {Suur-Uski}, {Tauber}, {Tavagnacco}, {Tenti}, {Toffolatti},
  {Tomasi}, {Trombetti}, {Valenziano}, {Valiviita}, {Van Tent}, {Vibert},
  {Vielva}, {Villa}, {Vittorio}, {Wandelt}, {Wehus}, {White}, {White},
  {Zacchei}, \& {Zonca}}]{planck20}
{Planck Collaboration}, {Aghanim}, N., {Akrami}, Y., {et~al.} 2020, \aap, 641,
  A6

\bibitem[{{Price-Whelan} {et~al.}(2018){Price-Whelan}, {Sip{\H{o}}cz},
  {G{\"u}nther}, {Lim}, {Crawford}, {Conseil}, {Shupe}, {Craig}, {Dencheva},
  {Ginsburg}, {VanderPlas}, {Bradley}, {P{\'e}rez-Su{\'a}rez}, {de Val-Borro},
  {Paper Contributors}, {Aldcroft}, {Cruz}, {Robitaille}, {Tollerud},
  {Coordination Committee}, {Ardelean}, {Babej}, {Bach}, {Bachetti}, {Bakanov},
  {Bamford}, {Barentsen}, {Barmby}, {Baumbach}, {Berry}, {Biscani}, {Boquien},
  {Bostroem}, {Bouma}, {Brammer}, {Bray}, {Breytenbach}, {Buddelmeijer},
  {Burke}, {Calderone}, {Cano Rodr{\'\i}guez}, {Cara}, {Cardoso}, {Cheedella},
  {Copin}, {Corrales}, {Crichton}, {D{\textquoteright}Avella}, {Deil},
  {Depagne}, {Dietrich}, {Donath}, {Droettboom}, {Earl}, {Erben}, {Fabbro},
  {Ferreira}, {Finethy}, {Fox}, {Garrison}, {Gibbons}, {Goldstein}, {Gommers},
  {Greco}, {Greenfield}, {Groener}, {Grollier}, {Hagen}, {Hirst}, {Homeier},
  {Horton}, {Hosseinzadeh}, {Hu}, {Hunkeler}, {Ivezi{\'c}}, {Jain}, {Jenness},
  {Kanarek}, {Kendrew}, {Kern}, {Kerzendorf}, {Khvalko}, {King}, {Kirkby},
  {Kulkarni}, {Kumar}, {Lee}, {Lenz}, {Littlefair}, {Ma}, {Macleod},
  {Mastropietro}, {McCully}, {Montagnac}, {Morris}, {Mueller}, {Mumford},
  {Muna}, {Murphy}, {Nelson}, {Nguyen}, {Ninan}, {N{\"o}the}, {Ogaz}, {Oh},
  {Parejko}, {Parley}, {Pascual}, {Patil}, {Patil}, {Plunkett}, {Prochaska},
  {Rastogi}, {Reddy Janga}, {Sabater}, {Sakurikar}, {Seifert}, {Sherbert},
  {Sherwood-Taylor}, {Shih}, {Sick}, {Silbiger}, {Singanamalla}, {Singer},
  {Sladen}, {Sooley}, {Sornarajah}, {Streicher}, {Teuben}, {Thomas},
  {Tremblay}, {Turner}, {Terr{\'o}n}, {van Kerkwijk}, {de la Vega}, {Watkins},
  {Weaver}, {Whitmore}, {Woillez}, {Zabalza}, \& {Contributors}}]{astropy2}
{Price-Whelan}, A.~M., {Sip{\H{o}}cz}, B.~M., {G{\"u}nther}, H.~M., {et~al.}
  2018, \aj, 156, 123

\bibitem[{{Refsdal}(1964)}]{refsdal64}
{Refsdal}, S. 1964, \mnras, 128, 307

\bibitem[{{Rodney} {et~al.}(2021){Rodney}, {Brammer}, {Pierel}, {Richard},
  {Toft}, {O'Connor}, {Akhshik}, \& {Whitaker}}]{rodney21}
{Rodney}, S.~A., {Brammer}, G.~B., {Pierel}, J. D.~R., {et~al.} 2021, Nature
  Astronomy, 5, 1118

\bibitem[{{Rojas} {et~al.}(2023){Rojas}, {Collett}, {Ballard}, {Magee},
  {Birrer}, {Buckley-Geer}, {Chan}, {Cl{\'e}ment}, {Diego}, {Gentile},
  {Gonz{\'a}lez}, {Joseph}, {Mastache}, {Schuldt}, {Tortora}, {Verdugo},
  {Verma}, {Daylan}, {Millon}, {Jackson}, {Dye}, {Melo}, {Mahler}, {Ogando},
  {Courbin}, {Fritz}, {Herle}, {Acevedo Barroso}, {Ca{\~n}ameras}, {Cornen},
  {Dhanasingham}, {Glazebrook}, {Martinez}, {Ryczanowski}, {Savary},
  {G{\'o}is-Silva}, {Arturo Ure{\~n}a-L{\'o}pez}, {Wiesner}, {Wilde}, {Valim
  Cal{\c{c}}ada}, {Cabanac}, {Pan}, {Sierra}, {Despali}, {Cavalcante-Gomes},
  {Macmillan}, {Maresca}, {Grudskaia}, {O'Donnell}, {Paic}, {Niemiec}, {de la
  Bella}, {Bromley}, {Williams}, {More}, \& {Levine}}]{rojas23}
{Rojas}, K., {Collett}, T.~E., {Ballard}, D., {et~al.} 2023, \mnras, 523, 4413

\bibitem[{{Rojas} {et~al.}(2022){Rojas}, {Savary}, {Cl{\'e}ment}, {Maus},
  {Courbin}, {Lemon}, {Chan}, {Vernardos}, {Joseph}, {Ca{\~n}ameras}, \&
  {Galan}}]{rojas22}
{Rojas}, K., {Savary}, E., {Cl{\'e}ment}, B., {et~al.} 2022, \aap, 668, A73

\bibitem[{{Savary} {et~al.}(2022){Savary}, {Rojas}, {Maus}, {Cl{\'e}ment},
  {Courbin}, {Gavazzi}, {Chan}, {Lemon}, {Vernardos}, {Ca{\~n}ameras},
  {Schuldt}, {Suyu}, {Cuillandre}, {Fabbro}, {Gwyn}, {Hudson}, {Kilbinger},
  {Scott}, \& {Stone}}]{savary22}
{Savary}, E., {Rojas}, K., {Maus}, M., {et~al.} 2022, \aap, 666, A1

\bibitem[{{Schuldt} {et~al.}(2025){Schuldt}, {Ca{\~n}ameras}, {Andika}, {Bag},
  {Melo}, {Shu}, {Suyu}, {Taubenberger}, \& {Grillo}}]{schuldt25a}
{Schuldt}, S., {Ca{\~n}ameras}, R., {Andika}, I.~T., {et~al.} 2025, \aap, 693,
  A291

\bibitem[{{Schuldt} {et~al.}(2023{\natexlab{a}}){Schuldt}, {Ca{\~n}ameras},
  {Shu}, {Suyu}, {Taubenberger}, {Meinhardt}, \& {Leal-Taix{\'e}}}]{schuldt23a}
{Schuldt}, S., {Ca{\~n}ameras}, R., {Shu}, Y., {et~al.} 2023{\natexlab{a}},
  \aap, 671, A147

\bibitem[{{Schuldt} {et~al.}(2019){Schuldt}, {Chiriv{\`\i}}, {Suyu},
  {Y{\i}ld{\i}r{\i}m}, {Sonnenfeld}, {Halkola}, \& {Lewis}}]{schuldt19}
{Schuldt}, S., {Chiriv{\`\i}}, G., {Suyu}, S.~H., {et~al.} 2019, \aap, 631, A40

\bibitem[{{Schuldt} {et~al.}(2023{\natexlab{b}}){Schuldt}, {Suyu},
  {Ca{\~n}ameras}, {Shu}, {Taubenberger}, {Ertl}, \& {Halkola}}]{schuldt23b}
{Schuldt}, S., {Suyu}, S.~H., {Ca{\~n}ameras}, R., {et~al.} 2023{\natexlab{b}},
  \aap, 673, A33

\bibitem[{{Schuldt} {et~al.}(2021){Schuldt}, {Suyu}, {Ca{\~n}ameras},
  {Taubenberger}, {Meinhardt}, {Leal-Taix{\'e}}, \& {Hsieh}}]{schuldt21b}
{Schuldt}, S., {Suyu}, S.~H., {Ca{\~n}ameras}, R., {et~al.} 2021, \aap, 651,
  A55

\bibitem[{{Shajib} {et~al.}(2021){Shajib}, {Treu}, {Birrer}, \&
  {Sonnenfeld}}]{shajib21}
{Shajib}, A.~J., {Treu}, T., {Birrer}, S., \& {Sonnenfeld}, A. 2021, \mnras,
  503, 2380

\bibitem[{{Shajib} {et~al.}(2022){Shajib}, {Wong}, {Birrer}, {Suyu}, {Treu},
  {Buckley-Geer}, {Lin}, {Rusu}, {Poh}, {Palmese}, {Agnello}, {Auger-Williams},
  {Galan}, {Schuldt}, {Sluse}, {Courbin}, {Frieman}, \& {Millon}}]{shajib22}
{Shajib}, A.~J., {Wong}, K.~C., {Birrer}, S., {et~al.} 2022, \aap, 667, A123

\bibitem[{{Shu} {et~al.}(2016){Shu}, {Bolton}, {Kochanek}, {Oguri},
  {P{\'e}rez-Fournon}, {Zheng}, {Mao}, {Montero-Dorta}, {Brownstein},
  {Marques-Chaves}, \& {M{\'e}nard}}]{shu16a_BELLS_search}
{Shu}, Y., {Bolton}, A.~S., {Kochanek}, C.~S., {et~al.} 2016, \apj, 824, 86

\bibitem[{{Shu} {et~al.}(2017){Shu}, {Brownstein}, {Bolton}, {Koopmans},
  {Treu}, {Montero-Dorta}, {Auger}, {Czoske}, {Gavazzi}, {Marshall}, \&
  {Moustakas}}]{shu17}
{Shu}, Y., {Brownstein}, J.~R., {Bolton}, A.~S., {et~al.} 2017, \apj, 851, 48

\bibitem[{{Shu} {et~al.}(2022){Shu}, {Ca{\~n}ameras}, {Schuldt}, {Suyu},
  {Taubenberger}, {Inoue}, \& {Jaelani}}]{shu22}
{Shu}, Y., {Ca{\~n}ameras}, R., {Schuldt}, S., {et~al.} 2022, \aap, 662, A4

\bibitem[{{Shu} {et~al.}(2018){Shu}, {Marques-Chaves}, {Evans}, \&
  {P{\'e}rez-Fournon}}]{shu18}
{Shu}, Y., {Marques-Chaves}, R., {Evans}, N.~W., \& {P{\'e}rez-Fournon}, I.
  2018, \mnras, 481, L136

\bibitem[{{Sonnenfeld} {et~al.}(2018){Sonnenfeld}, {Chan}, {Shu}, {More},
  {Oguri}, {Suyu}, {Wong}, {Lee}, {Coupon}, {Yonehara}, {Bolton}, {Jaelani},
  {Tanaka}, {Miyazaki}, \& {Komiyama}}]{sonnenfeld18a}
{Sonnenfeld}, A., {Chan}, J. H.~H., {Shu}, Y., {et~al.} 2018, \pasj, 70, S29

\bibitem[{{Sonnenfeld} {et~al.}(2020){Sonnenfeld}, {Verma}, {More}, {Baeten},
  {Macmillan}, {Wong}, {Chan}, {Jaelani}, {Lee}, {Oguri}, {Rusu}, {Veldthuis},
  {Trouille}, {Marshall}, {Hutchings}, {Allen}, {O'Donnell}, {Cornen}, {Davis},
  {McMaster}, {Lintott}, \& {Miller}}]{sonnenfeld20}
{Sonnenfeld}, A., {Verma}, A., {More}, A., {et~al.} 2020, \aap, 642, A148

\bibitem[{{Spergel} {et~al.}(2015){Spergel}, {Gehrels}, {Baltay}, {Bennett},
  {Breckinridge}, {Donahue}, {Dressler}, {Gaudi}, {Greene}, {Guyon}, {Hirata},
  {Kalirai}, {Kasdin}, {Macintosh}, {Moos}, {Perlmutter}, {Postman},
  {Rauscher}, {Rhodes}, {Wang}, {Weinberg}, {Benford}, {Hudson}, {Jeong},
  {Mellier}, {Traub}, {Yamada}, {Capak}, {Colbert}, {Masters}, {Penny},
  {Savransky}, {Stern}, {Zimmerman}, {Barry}, {Bartusek}, {Carpenter}, {Cheng},
  {Content}, {Dekens}, {Demers}, {Grady}, {Jackson}, {Kuan}, {Kruk}, {Melton},
  {Nemati}, {Parvin}, {Poberezhskiy}, {Peddie}, {Ruffa}, {Wallace}, {Whipple},
  {Wollack}, \& {Zhao}}]{spergel15_roman}
{Spergel}, D., {Gehrels}, N., {Baltay}, C., {et~al.} 2015, arXiv e-prints,
  arXiv:1503.03757

\bibitem[{{Stein} {et~al.}(2022){Stein}, {Blaum}, {Harrington}, {Medan}, \&
  {Luki{\'c}}}]{stein22}
{Stein}, G., {Blaum}, J., {Harrington}, P., {Medan}, T., \& {Luki{\'c}}, Z.
  2022, \apj, 932, 107

\bibitem[{{Stiavelli} {et~al.}(2023){Stiavelli}, {Morishita}, {Chiaberge},
  {Grillo}, {Leethochawalit}, {Rosati}, {Schuldt}, {Trenti}, \&
  {Treu}}]{stiavelli23}
{Stiavelli}, M., {Morishita}, T., {Chiaberge}, M., {et~al.} 2023, \apjl, 957,
  L18

\bibitem[{{Storfer} {et~al.}(2024){Storfer}, {Huang}, {Gu}, {Sheu}, {Banka},
  {Dey}, {Inchausti Reyes}, {Jain}, {Kwon}, {Lang}, {Lee}, {Meisner},
  {Moustakas}, {Myers}, {Tabares-Tarquinio}, {Schlafly}, \&
  {Schlegel}}]{storfer24}
{Storfer}, C., {Huang}, X., {Gu}, A., {et~al.} 2024, \apjs, 274, 16

\bibitem[{{Suyu} \& {Halkola}(2010)}]{suyu10a_GLEE}
{Suyu}, S.~H. \& {Halkola}, A. 2010, \aap, 524, A94

\bibitem[{{Suyu} {et~al.}(2012{\natexlab{a}}){Suyu}, {Hensel}, {McKean},
  {Fassnacht}, {Treu}, {Halkola}, {Norbury}, {Jackson}, {Schneider},
  {Thompson}, {Auger}, {Koopmans}, \& {Matthews}}]{suyu12b_GLEE}
{Suyu}, S.~H., {Hensel}, S.~W., {McKean}, J.~P., {et~al.} 2012{\natexlab{a}},
  \apj, 750, 10

\bibitem[{{Suyu} {et~al.}(2020){Suyu}, {Huber}, {Ca{\~n}ameras}, {Kromer},
  {Schuldt}, {Taubenberger}, {Y{\i}ld{\i}r{\i}m}, {Bonvin}, {Chan}, {Courbin},
  {N{\"o}bauer}, {Sim}, \& {Sluse}}]{suyu20}
{Suyu}, S.~H., {Huber}, S., {Ca{\~n}ameras}, R., {et~al.} 2020, \aap, 644, A162

\bibitem[{{Suyu} {et~al.}(2012{\natexlab{b}}){Suyu}, {Treu}, {Blandford},
  {Freedman}, {Hilbert}, {Blake}, {Braatz}, {Courbin}, {Dunkley}, {Greenhill},
  {Humphreys}, {Jha}, {Kirshner}, {Lo}, {Macri}, {Madore}, {Marshall},
  {Meylan}, {Mould}, {Reid}, {Reid}, {Riess}, {Schlegel}, {Scowcroft}, \&
  {Verde}}]{suyu12}
{Suyu}, S.~H., {Treu}, T., {Blandford}, R.~D., {et~al.} 2012{\natexlab{b}},
  arXiv e-prints, arXiv:1202.4459

\bibitem[{{Talbot} {et~al.}(2021){Talbot}, {Brownstein}, {Dawson}, {Kneib}, \&
  {Bautista}}]{talbot21}
{Talbot}, M.~S., {Brownstein}, J.~R., {Dawson}, K.~S., {Kneib}, J.-P., \&
  {Bautista}, J. 2021, \mnras, 502, 4617

\bibitem[{{Tanaka} {et~al.}(2018){Tanaka}, {Coupon}, {Hsieh}, {Mineo},
  {Nishizawa}, {Speagle}, {Furusawa}, {Miyazaki}, \& {Murayama}}]{tanaka18}
{Tanaka}, M., {Coupon}, J., {Hsieh}, B.-C., {et~al.} 2018, \pasj, 70, S9

\bibitem[{{van der Walt} {et~al.}(2011){van der Walt}, {Colbert}, \&
  {Varoquaux}}]{numpy1}
{van der Walt}, S., {Colbert}, S.~C., \& {Varoquaux}, G. 2011, Computing in
  Science Engineering, 13, 22

\bibitem[{Van~Rossum \& Drake(2009)}]{python}
Van~Rossum, G. \& Drake, F.~L. 2009, Python 3 Reference Manual (Scotts Valley,
  CA: CreateSpace)

\bibitem[{{Vanzella} {et~al.}(2021){Vanzella}, {Caminha}, {Rosati}, {Mercurio},
  {Castellano}, {Meneghetti}, {Grillo}, {Sani}, {Bergamini}, {Calura},
  {Caputi}, {Cristiani}, {Cupani}, {Fontana}, {Gilli}, {Grazian}, {Gronke},
  {Mignoli}, {Nonino}, {Pentericci}, {Tozzi}, {Treu}, {Balestra}, \&
  {Dijkstra}}]{vanzella21}
{Vanzella}, E., {Caminha}, G.~B., {Rosati}, P., {et~al.} 2021, \aap, 646, A57

\bibitem[{{Virtanen} {et~al.}(2020){Virtanen}, {Gommers}, {Oliphant},
  {Haberland}, {Reddy}, {Cournapeau}, {Burovski}, {Peterson}, {Weckesser},
  {Bright}, {van der Walt}, {Brett}, {Wilson}, {Jarrod Millman}, {Mayorov},
  {Nelson}, {Jones}, {Kern}, {Larson}, {Carey}, {Polat}, {Feng}, {Moore}, {Vand
  erPlas}, {Laxalde}, {Perktold}, {Cimrman}, {Henriksen}, {Quintero}, {Harris},
  {Archibald}, {Ribeiro}, {Pedregosa}, {van Mulbregt}, \&
  {Contributors}}]{scipy}
{Virtanen}, P., {Gommers}, R., {Oliphant}, T.~E., {et~al.} 2020, Nature
  Methods, 17, 261

\bibitem[{{Wang} {et~al.}(2022){Wang}, {Ca{\~n}ameras}, {Caminha}, {Suyu},
  {Y{\i}ld{\i}r{\i}m}, {Chiriv{\`\i}}, {Christensen}, {Grillo}, \&
  {Schuldt}}]{wang22}
{Wang}, H., {Ca{\~n}ameras}, R., {Caminha}, G.~B., {et~al.} 2022, \aap, 668,
  A162

\bibitem[{{Wong} {et~al.}(2022){Wong}, {Chan}, {Chao}, {Jaelani}, {Kayo},
  {Lee}, {More}, \& {Oguri}}]{wong22}
{Wong}, K.~C., {Chan}, J. H.~H., {Chao}, D. C.~Y., {et~al.} 2022, \pasj, 74,
  1209

\bibitem[{{Wong} {et~al.}(2018){Wong}, {Sonnenfeld}, {Chan}, {Rusu}, {Tanaka},
  {Jaelani}, {Lee}, {More}, {Oguri}, {Suyu}, \& {Komiyama}}]{wong18}
{Wong}, K.~C., {Sonnenfeld}, A., {Chan}, J. H.~H., {et~al.} 2018, \apj, 867,
  107

\bibitem[{{Wong} {et~al.}(2020){Wong}, {Suyu}, {Chen}, {Rusu}, {Millon},
  {Sluse}, {Bonvin}, {Fassnacht}, {Taubenberger}, {Auger}, {Birrer}, {Chan},
  {Courbin}, {Hilbert}, {Tihhonova}, {Treu}, {Agnello}, {Ding}, {Jee},
  {Komatsu}, {Shajib}, {Sonnenfeld}, {Blandford}, {Koopmans}, {Marshall}, \&
  {Meylan}}]{wong20}
{Wong}, K.~C., {Suyu}, S.~H., {Chen}, G. C.~F., {et~al.} 2020, \mnras, 498,
  1420

\bibitem[{{Zhong} {et~al.}(2022){Zhong}, {Li}, \& {Napolitano}}]{zhong22}
{Zhong}, F., {Li}, R., \& {Napolitano}, N.~R. 2022, Research in Astronomy and
  Astrophysics, 22, 065014

\end{thebibliography}

\newpage

\onecolumn
\appendix

\section{Detailed description of the released catalog of inspected lens candidates}

With this publication, we have release the full catalog of inspected lens candidates. This catalog is introduced in Table~\ref{tab:newcand} and electronically available at the HOLISMOKES webpage\footnote{\url{www.holismokes.org}}, the SuGOHI database\footnote{\url{https://www-utap.phys.s.u-tokyo.ac.jp/˜oguri/sugohi/}}, and CDS. The columns are described in Table~\ref{tab:legend}.

\begin{table*}[h!]
    \caption{Detailed description of the released catalog, as previewed with five lines in Table~\ref{tab:newcand} for a selected subset of columns.}
    \begin{center}
    \begin{tabular}{ll|l}
  Column & Header & Description \\   \hline \hline
  (1) & Name & name for grade A and B lens candidates  \\ \hline
  (2) & RA [deg] & Right Ascension of the inspected candidate\\ 
  (3) & Dec [deg] & Declination of the inspected candidate\\ \hline
  (4) & p & average score predicted by the network committee \\
  (5) & G & average grade between 0 and 3 obtained through visual inspection (see Sect.~\ref{sec:visualinspection})\\
  (6) & sigma\_G & standard deviation of obtained visual inspection grades\\
  (7) & N\_graders & number of visual inspectors (see Sect.~\ref{sec:visualinspection})\\
  (8) & $i$\_kronflux\_radius & Kron radius in the $i$ band provided by HSC used for source selection (see Sect.~\ref{sec:network})\\ \hline
  (9) & Binary & Flag to indicate sources inspected only in the binary stage (two different graders)\\
  (10) & Round1 & Flag to indicate systems inspected also in the first round (four different graders)\\
  (11) & Round2 & Flag to indicate systems inspected also in the second round (eight different graders)\\
  (12) & A24 & Flag to indicate if system jointly inspected with candidates discovered by \citet{andika25}\\
  (13) & S25 & Flag to indicate systems discovered by this network committee but not re-inspected. Instead, we report the \\
  & & visual inspection grade from our earlier work \citetalias{schuldt25a}\\
  (14) & S22 & Flag to indicate systems discovered by this network committee but not re-inspected. Instead, we report the \\
  & & visual inspection grade from our earlier work \citet{shu22}\\
  (15) & C21 & Flag to indicate systems discovered by this network committee but not re-inspected. Instead, we report the \\
  & & visual inspection grade from our earlier work \citet{canameras21b}\\ \hline
  (16) & x\_med & $x$-center coordinate predicted by the modeling network from \citet[][hereafter \citetalias{schuldt23a}]{schuldt23a}\\
  (17) & x\_err & 1 $\sigma$ value for the $x$-center coordinate predicted by the modeling network from \citetalias{schuldt23a}\\
  (18) & y\_med & $y$-center coordinate predicted by the modeling network from \citetalias{schuldt23a}\\
  (19) & y\_err & 1 $\sigma$ value for the $y$-center coordinate predicted by the modeling network from \citetalias{schuldt23a}\\
  (20) & ex\_med & $x$-component of the complex ellipticity predicted by the modeling network from \citetalias{schuldt23a}\\
  (21) & ex\_err & 1 $\sigma$ value for the $x$-component of the complex ellipticity predicted by the modeling network from \citetalias{schuldt23a}\\
  (22) & ey\_med & $y$-component of the complex ellipticity predicted by the modeling network from \citetalias{schuldt23a}\\
  (23) & ey\_err & 1 $\sigma$ value for the $y$-component of the complex ellipticity predicted by the modeling network from \citetalias{schuldt23a}\\
  (24) & rE\_med $[\arcsec]$ & Einstein radius value of the given candidate predicted by the modeling network from \citetalias{schuldt23a}\\
  (25) & rE\_err $[\arcsec]$ & 1 $\sigma$ value of the Einstein radius predicted by the modeling network from \citetalias{schuldt23a}\\
  (26) & gam1\_med & $\gamma_1$-component of the external shear predicted by the modeling network from \citetalias{schuldt23a}\\
  (27) & gma1\_err & 1 $\sigma$ value of the $\gamma_1$-component predicted by the modeling network from \citetalias{schuldt23a}\\
  (28) & gam2\_med & $\gamma_2$-component of the external shear predicted by the modeling network from \citetalias{schuldt23a}\\
  (29) & gam2\_err & 1 $\sigma$ value of the $\gamma_2$-component predicted by the modeling network from \citetalias{schuldt23a}\\
  (30) & Model & Flag if the model predict is reliable, only for systems inspected in the second round \\ \hline
  (31) & z & photometric redshift value from the catalog compiled by \citetalias{schuldt25a}\\
  & & based on DEmP \citep{hsieh14}, Mizuki \citep{tanaka18}, and NetZ \citep{schuldt21b}\\
  (32) & OD\_vis & Flag if the system falls into a significant overdense region\\
  (33) & OD\_z & Flag if the system is in a significant overdense environment according to the criteria defined by \citetalias{schuldt25a}\\
  (34) & N\_max & Peak of the photo-$z$ histogram following the procedure of \citetalias{schuldt25a}\\
  (35) & zlow & Lower bound of N\_max in the photo-$z$ histogram indicating the redshift of the overdensity\\
  (36) & Ntot & Sum of objects with photo-$z$ within a box of 200\arcsec on a side. \\
  (37) & References & List of publications that report the inspected candidate (within 5\arcsec) as lens candidate according to the\\ 
  & & HOLISMOKES \citet{suyu20} lens compilation with status of the publication, using 
B04 for \\ & & \citet{bolton04}, 
C07 for \citet{cabanac07}, 
B08 for \citet{bolton08}, 
G14 for \citet{gavazzi14}, \\ & &
H15 for \citet{holwerda15}, 
M16 for \citet{more16b}, 
P16 for \citet{paraficz16}, 
S16 for \\ & & \citet{shu16a_BELLS_search}, 
D17 for \citet{diehl17}, 
J17 for \citet{jacobs17}, 
S18 for \citet{sonnenfeld18a}, \\ & &
W18 for \citet{wong18}, 
J19 for \citet{jacobs19}, 
L19 for \citet{li19}, 
P19 for \citet{petrillo19b}, 
\\ & & H20 for \citet{huang20}, 
C20 for \citet{chan20}, 
Ca20 for \citet{canameras20},
Cao20 for \\ & &  \citet{cao20}, 
L20 for \citet{li20}, 
J20 for \citet{jaelani20b}, 
S20 for \citet{sonnenfeld20}, 
C21 \\ & & for \citet{canameras21b}, 
H21 for \citet{huang21}, 
L21 for \citet{li21c_search_KiDS}, 
T21 for \citet{talbot21}, \\ & & 
R22 for \citet{rojas22}, 
S22 for \citet{shu22}, 
Sa22 for \citet{savary22}, 
St22 \citet{stein22}, \\ & & 
W22 for \citet{wong22}, 
Z22 for \citet{zhong22}, 
A23 for \citet{andika23}, 
J24 for \\ & & \citet{jaelani24},
G24 for \citet{grespan24}, 
St24 for \citet{storfer24}, 
S25 for \citet{schuldt25a}, \\ & & 
ML for the master lens catalog at  \url{http://admin.masterlens.org}, 
and `Guoyou Sun’ corresponds to \\ & & candidates identified by an amateur astronomer, Guoyou Sun, through visual inspections of HSC cutouts (see \\ & & \url{http://sunguoyou.lamost.org/glc.html.}).
\\
\end{tabular}
    \end{center}
    \label{tab:legend}
\end{table*}

\FloatBarrier
\section{Color-composite images of rediscovered lens candidates}
\label{sec:appendixA}

In this section we present the color-composite image stamps of the lens candidates that obtained an average grade above 1.5 (i.e. grade A or B) during our visual inspection, but are already known in the literature. We note that the lens candidates discovered by \citetalias{canameras21b}, \citetalias{shu22}, and \citetalias{schuldt25a}, which we excluded before visual inspection to lower the amount, are not shown. 

\begin{figure*}[thb!]
\centering
\includegraphics[width=\textwidth]{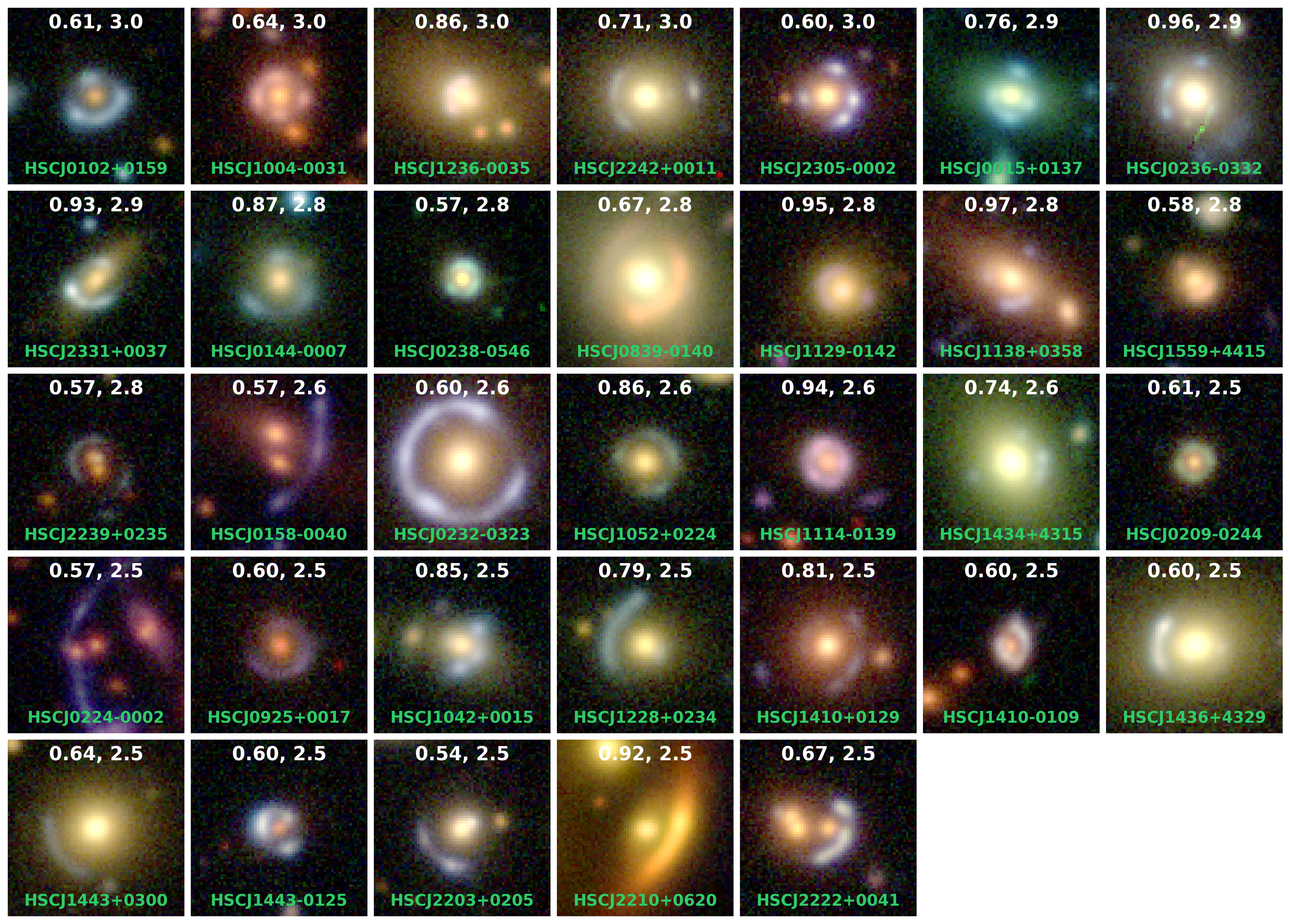}
\caption{Color-image stamps of rediscovered but again visually inspected grade-A lens candidates. Same format as Fig.~\ref{fig:gradeA}.}
\label{fig:gradeA_lit}
\end{figure*}

\begin{figure*}[thb!]
\centering
\includegraphics[width=0.91\textwidth]{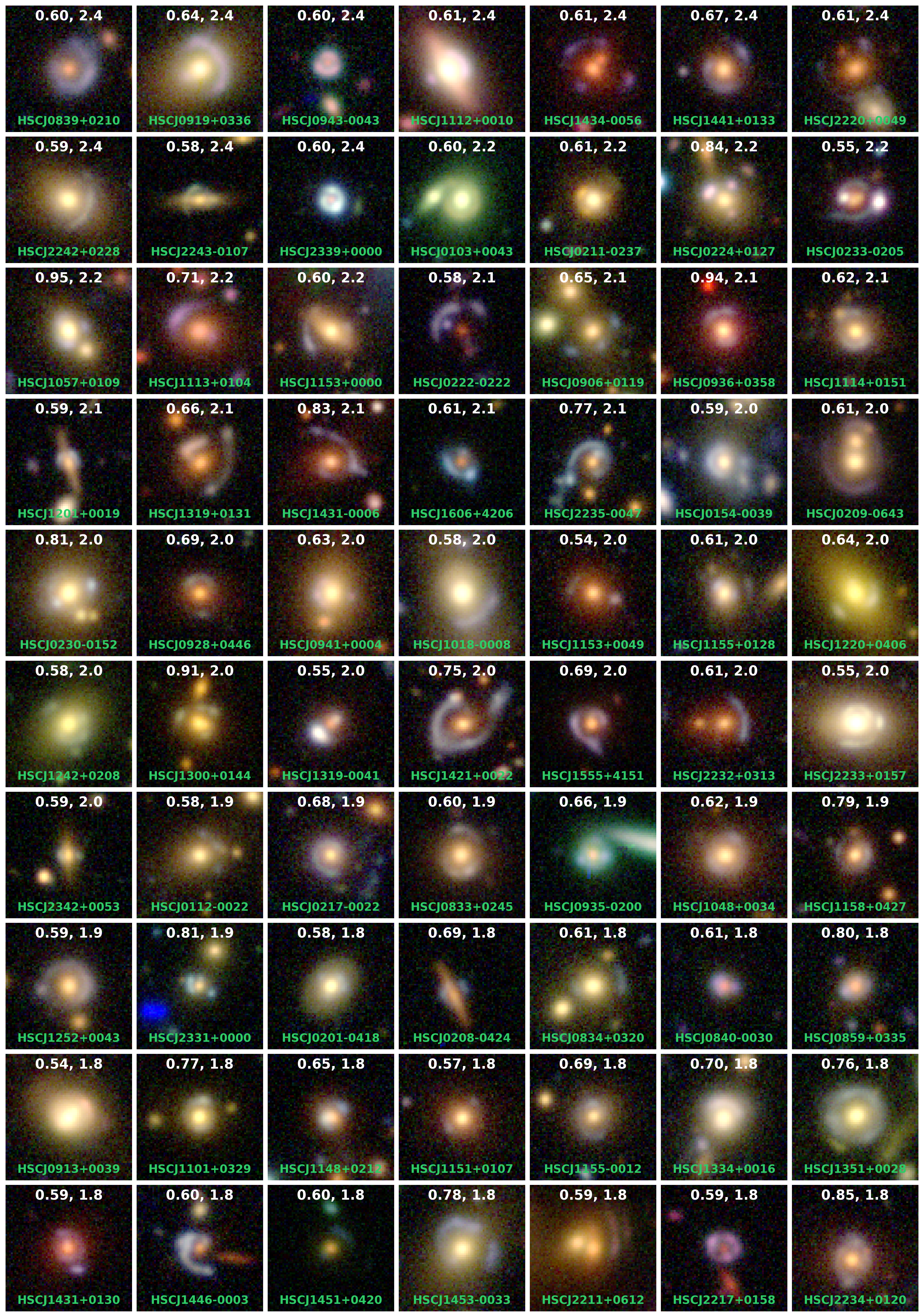}
\caption{Color-image stamps of rediscovered  but again visually inspected grade-B lens candidates. Same format as Fig.~\ref{fig:gradeA}.}
\label{fig:gradeB_lit}
\end{figure*}

\FloatBarrier

\begin{figure*}[thb!]
\centering
\includegraphics[width=\textwidth]{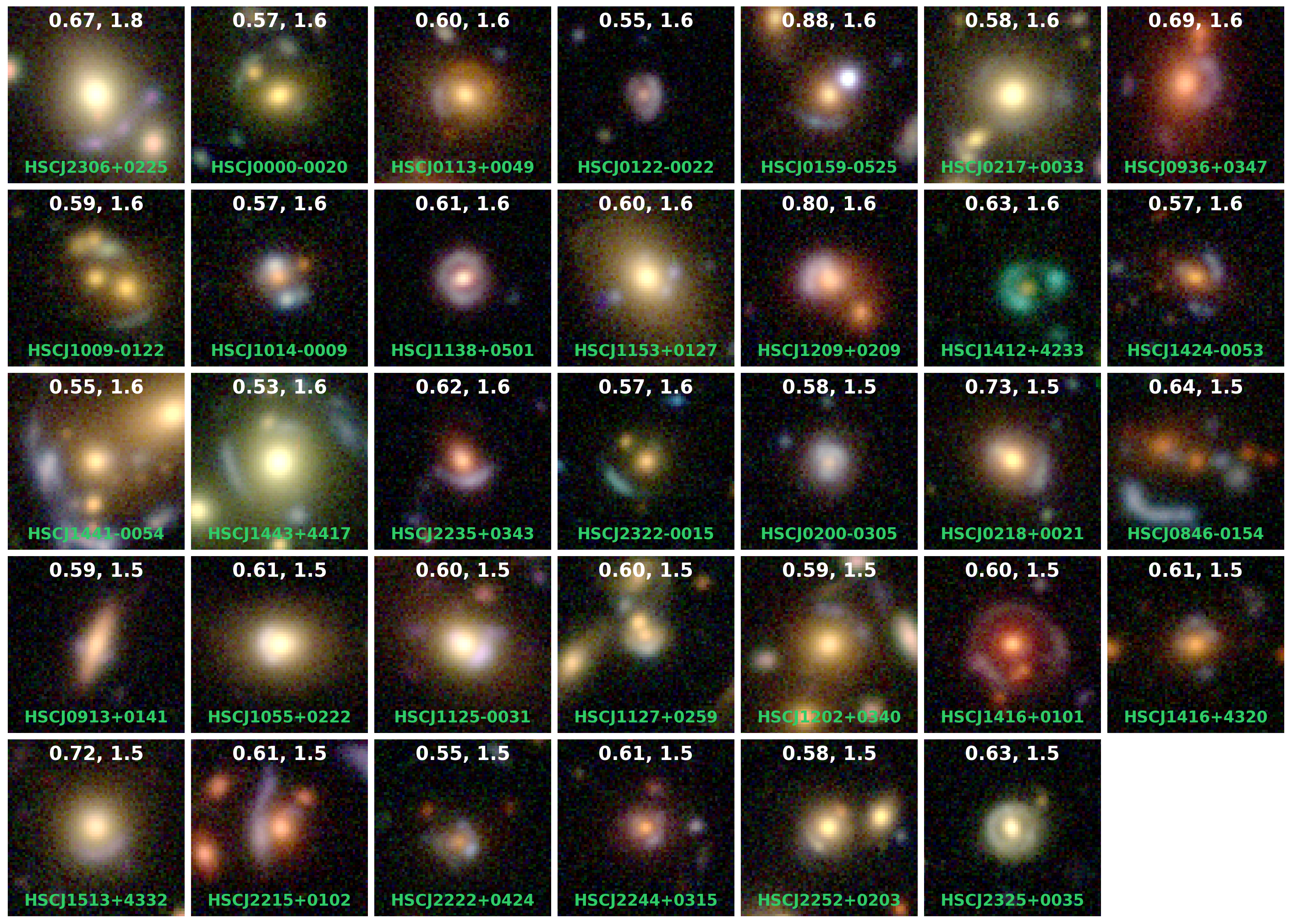}
\caption*{Fig.~\ref{fig:gradeB_lit} (continued): Color-image stamps of rediscovered but again visually inspected grade-B lens candidates. Same format as Fig.~\ref{fig:gradeA}.}
\label{fig:gradeB2_lit}
\end{figure*}

\section{Color-composite images of lens candidates with their predicted mass model}
\label{sec:appendixB}

In this section, we show the color-composite image stamps of the lens candidates in analogy to Figs.~\ref{fig:gradeA} and \ref{fig:gradeB}, as well as those in Sect.~\ref{sec:appendixA}. Contrary to previous figures, we show here the lens center and Einstein radius for each system predicted by the ResNet of \citet{schuldt23a}. These models were also shown in one panel during the visual inspection to help the grader classify.

\begin{figure*}[thb!]
\centering
\includegraphics[width=\textwidth]{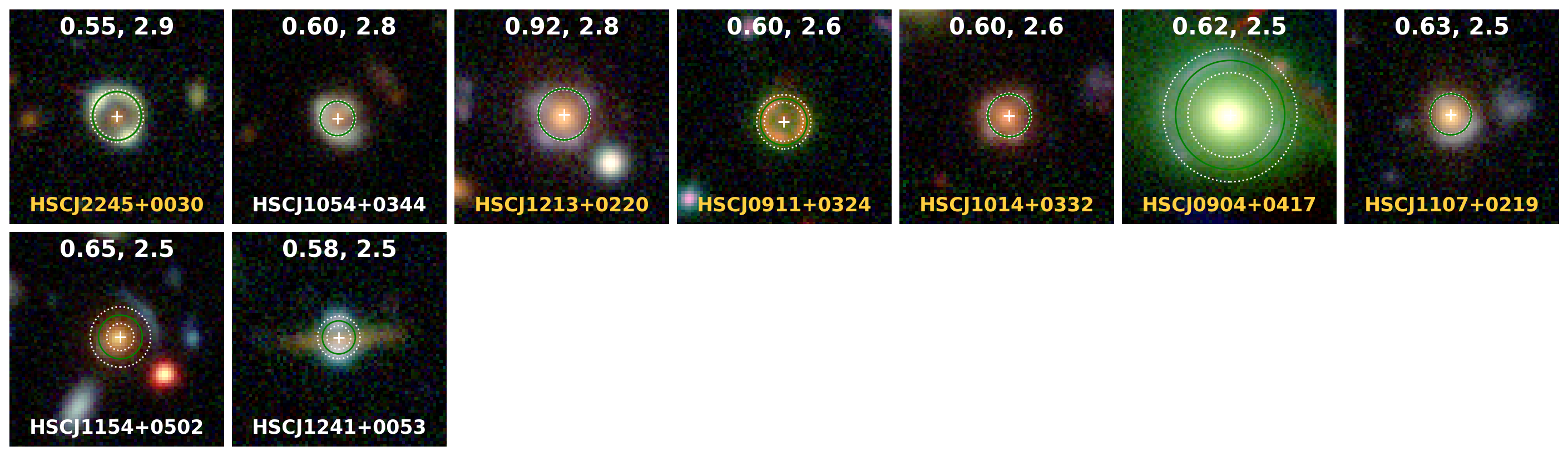}
\caption{Color-image stamps of newly discovered grade A lenses, but showing the lens center and the Einstein radius as green circles predicted by the ResNet of \citet{schuldt23a}. The predicted 1 $\sigma$ uncertainty for the Einstein radii are shown with white circles. We mark those with a well predicted lens center and Einstein radius (see also column 30 of Table~\ref{tab:newcand}) by yellow names instead of white. Remaining format as Fig.~\ref{fig:gradeA}.}
\label{fig:gradeA_model}
\end{figure*}

\begin{figure*}[thb!]
\centering
\includegraphics[width=\textwidth]{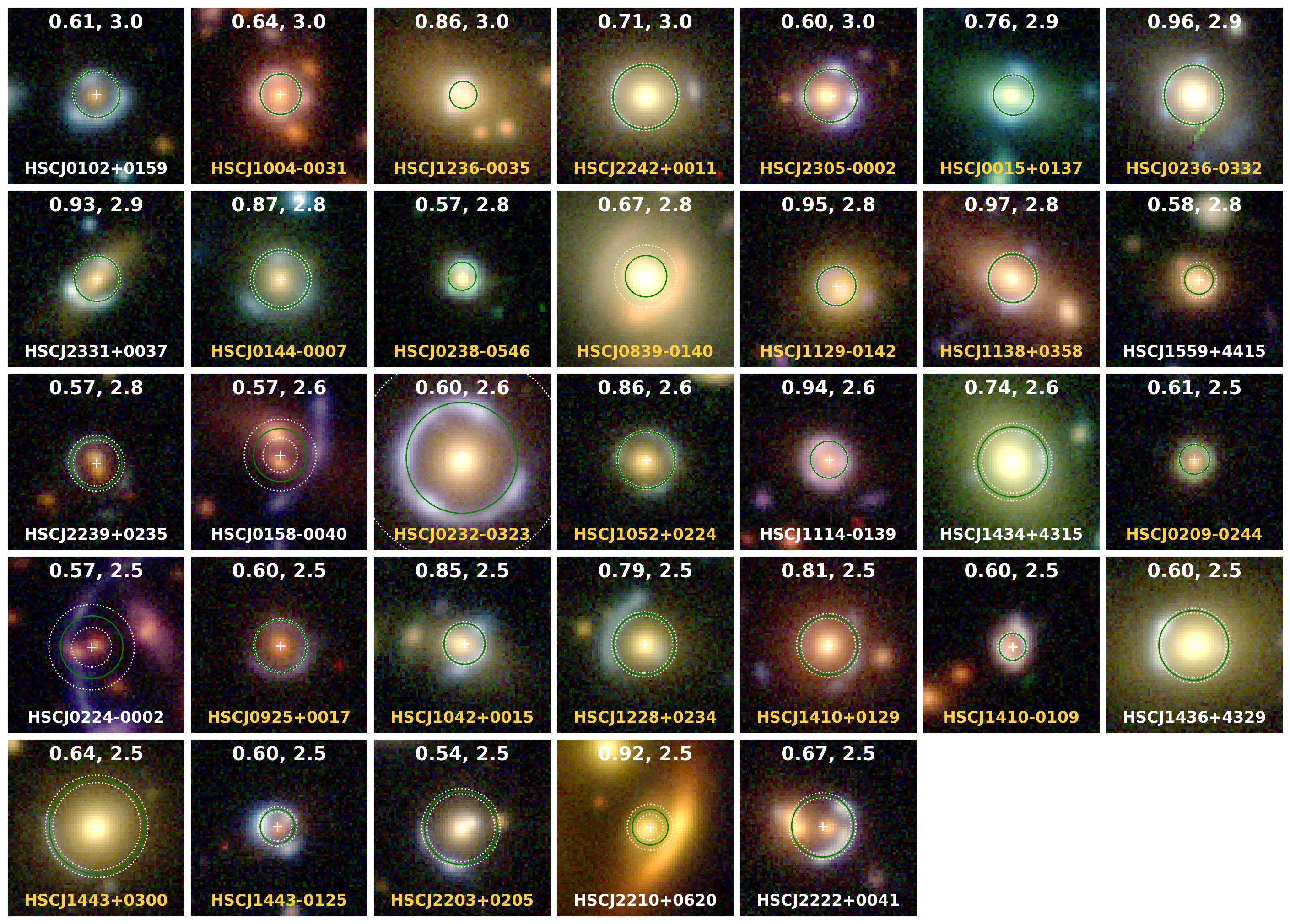}
\caption{Color-image stamps of rediscovered but visually re-inspected grade-A lens candidates. Same format as Fig.~\ref{fig:gradeA_model}.}
\label{fig:gradeA_lit_model}
\end{figure*}

\begin{figure*}[thb!]
\centering
\includegraphics[width=0.91\textwidth]{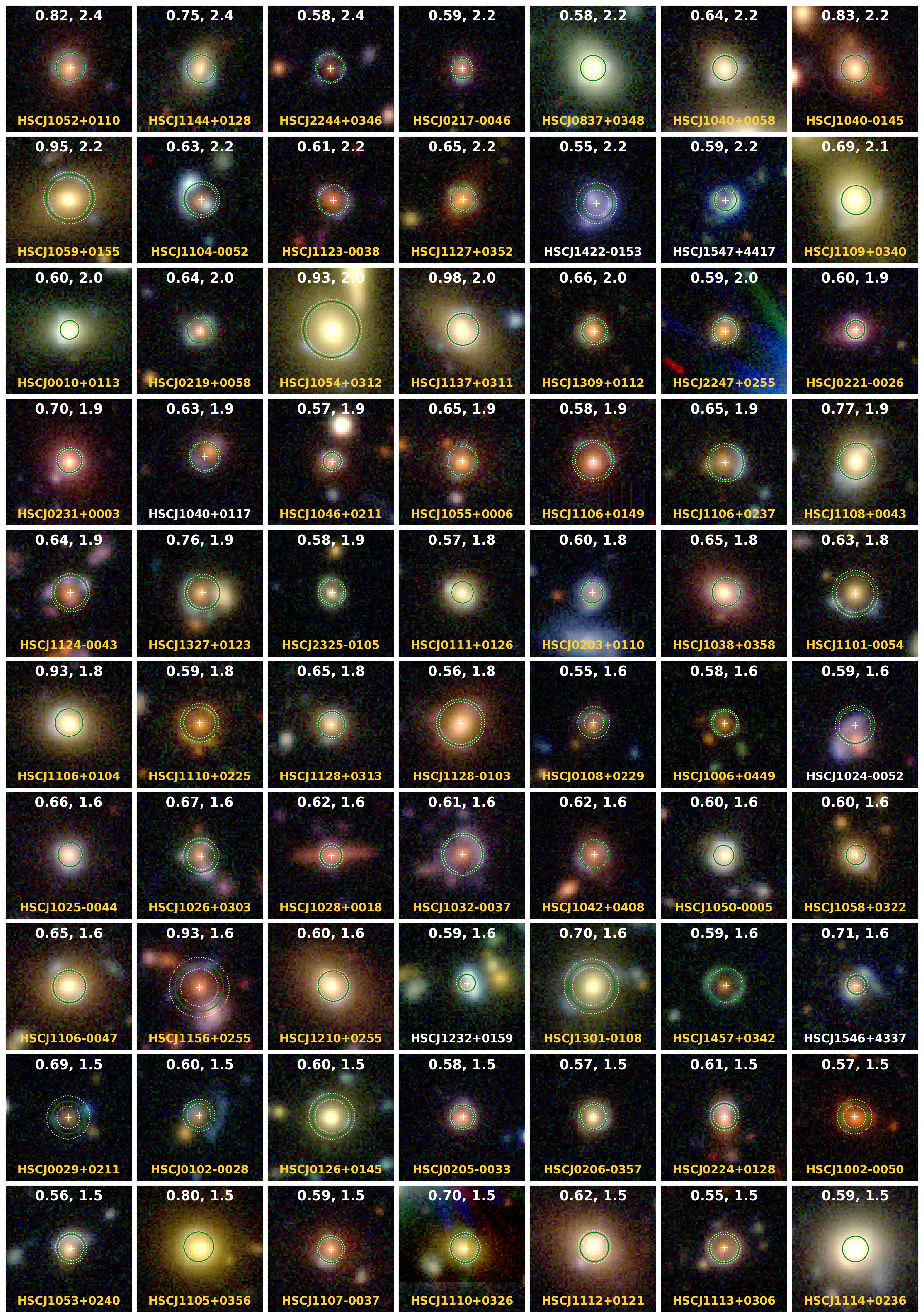}
\caption{Color-image stamps of newly discovered grade-B lens candidates. Same format as Fig.~\ref{fig:gradeA_model}.}
\label{fig:gradeB1_model}
\end{figure*}

\begin{figure*}[thb!]
\centering
\includegraphics[width=\textwidth]{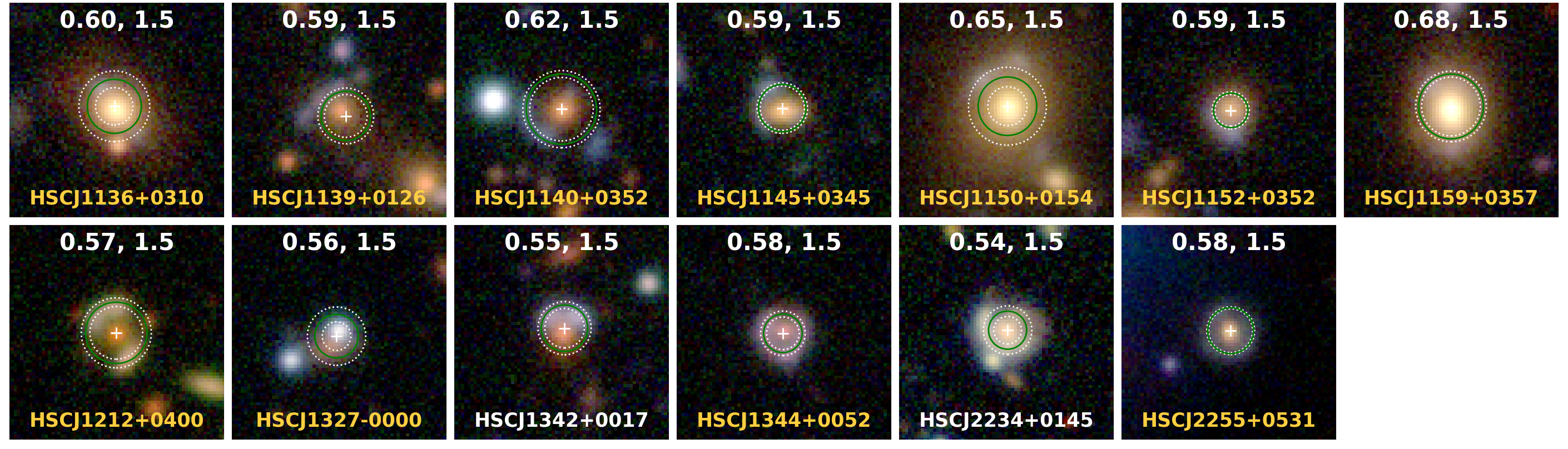}
\caption*{Fig.~\ref{fig:gradeB1_model} (continued): Color-image stamps of newly discovered grade-B lens candidates. Same format as Fig.~\ref{fig:gradeA_model}.}
\label{fig:gradeB2_model}
\end{figure*}

\begin{figure*}[thb!]
\centering
\includegraphics[width=0.91\textwidth]{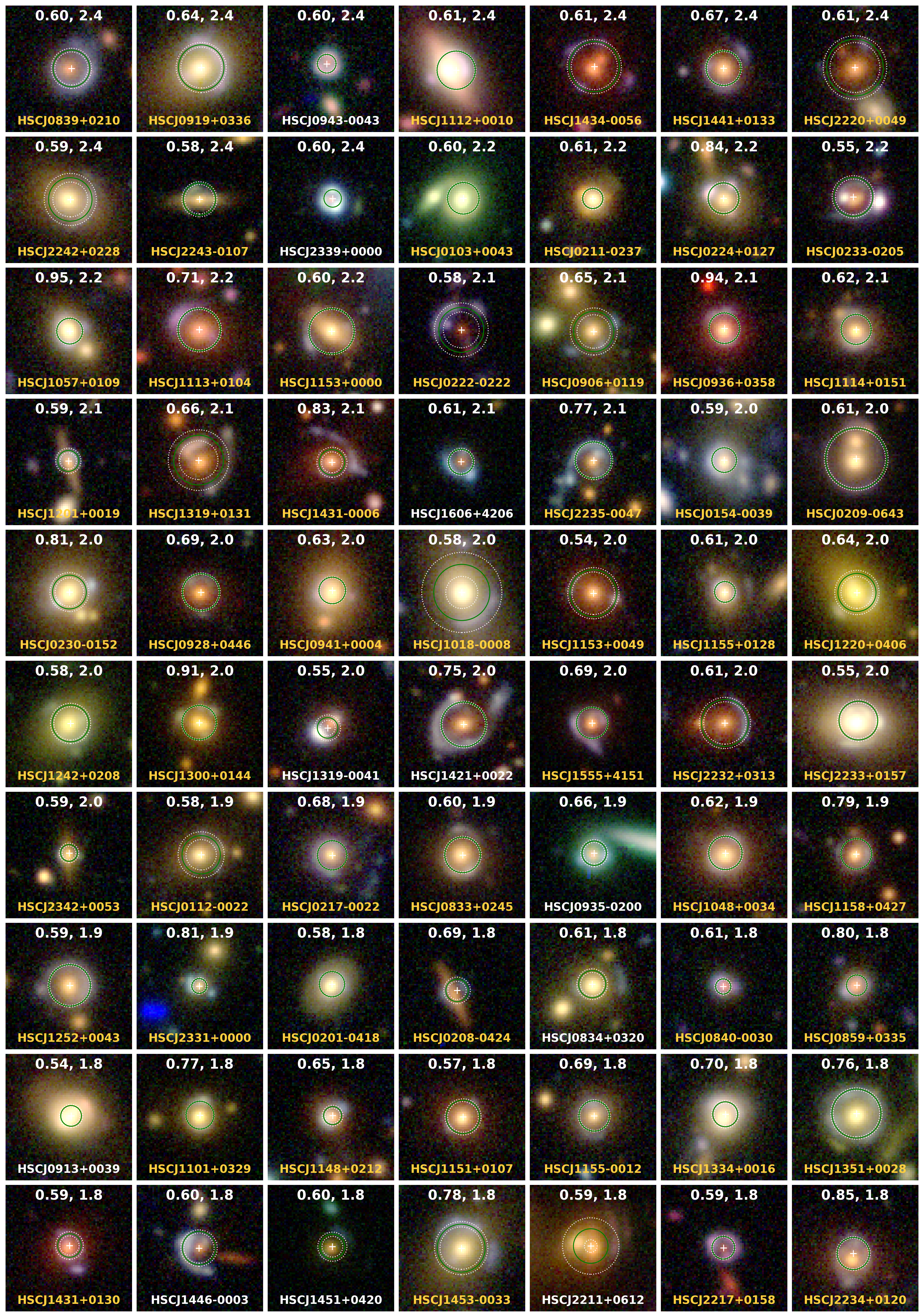}
\caption{Color-image stamps of rediscovered but visually re-inspected grade-B lens candidates. Same format as Fig.~\ref{fig:gradeA_model}.}
\label{fig:gradeB1_lit_model}
\end{figure*}

\FloatBarrier

\begin{figure*}[thb!]
\centering
\includegraphics[width=0.91\textwidth]{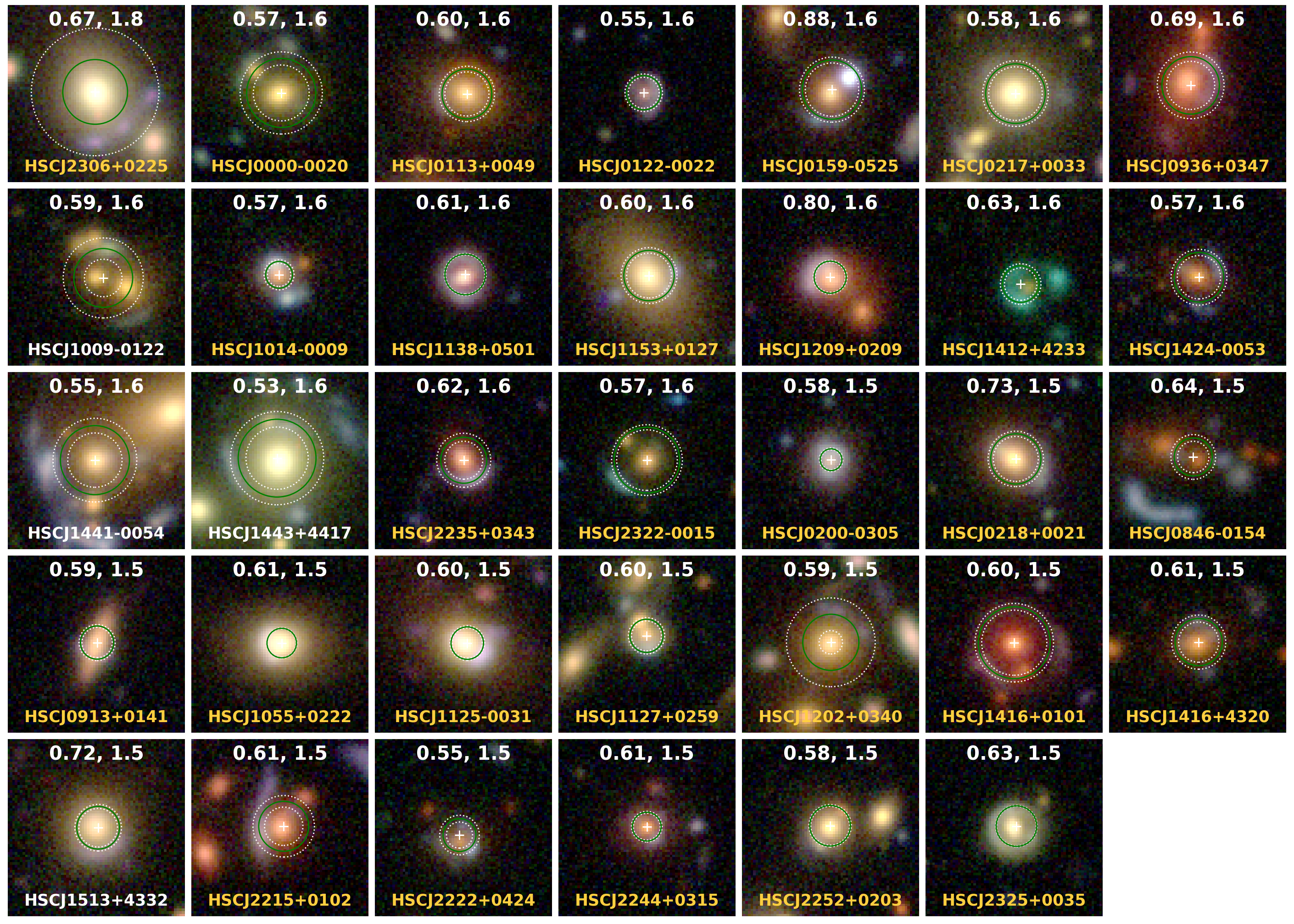}
\caption*{Fig.~\ref{fig:gradeB1_lit_model} (continued): Color-image stamps of rediscovered but visually re-inspected grade-B lens candidates. Format as Fig.~\ref{fig:gradeA_model}.}
\label{fig:gradeB2_lit_model}
\end{figure*}

\section{Comparison between inspected images from \citetalias{schuldt25a} and this work}

In this section, we show a representative sample of lens candidates that were graded in this work (see Sect.~\ref{sec:visualinspection}) and in \citetalias{schuldt25a}. While we showed during grading three different scalings, we show here only one filter combination for simplicity.

\begin{figure*}[thb!]
\centering
\includegraphics[trim= 15 250 130 110, clip, width=\textwidth]{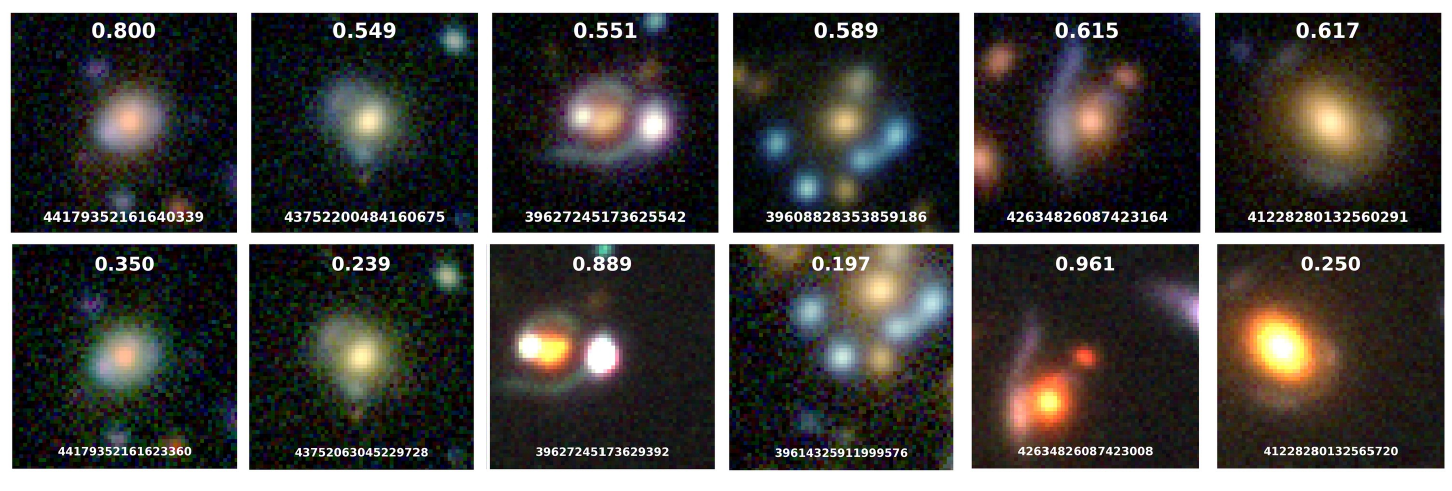}
\caption{Color-images ($gri$) shown for visual inspection in this work (top row) and in \citetalias{schuldt25a} (bottom row) to visualize the minor difference between PDR3 (this work) and PDR2 \citepalias{schuldt25a} and differences in the scaling of the individual filters because of the different cutout centering. We further show the network score (top of each panel) and the HSC ID (bottom).}
\label{fig:PDR3vsPDR2}
\end{figure*}

\end{document}